\documentclass[prd,aps,showpacs,nofootinbib,preprint,eqsecnum]{revtex4}
\usepackage[latin1]{inputenc}
\usepackage[T1]{fontenc}
\usepackage{amsmath}
\usepackage{amsfonts}
\usepackage{graphicx}
\usepackage{subfigure}
\usepackage{amssymb}
\usepackage{fancyhdr}
\usepackage{mathrsfs}
\usepackage{amsxtra}
\usepackage{epsf}
\usepackage{enumerate}
\usepackage{hhline}
\usepackage{array}
\usepackage{tabularx}
\usepackage[english]{babel}
\newcommand{\Eqn}[1]{&\hspace{-0.2em}#1\hspace{-0.2em}&}

\begin{document}

\title{Finite-time future singularities in modified Gauss-Bonnet and
$\mathcal{F}(R,G)$ gravity and singularity avoidance}

\author{Kazuharu Bamba$^{1,}$\footnote{E-mail address:
bamba@phys.nthu.edu.tw},
Sergei D. Odintsov$^{2,}$\footnote{Also at Tomsk State
Pedagogical University. E-mail address: odintsov@aliga.ieec.uab.es},
Lorenzo Sebastiani$^{3,}$\footnote{E-mail address:
l.sebastiani@science.unitn.it
}
and
Sergio Zerbini$^{3,}$\footnote{E-mail address:
zerbini@science.unitn.it}
}
\affiliation{
$^1$Department of Physics, National Tsing Hua University, Hsinchu, Taiwan
300\\
$^2$Instituci\`{o} Catalana de Recerca i Estudis Avan\c{c}ats (ICREA)
and Institut de Ciencies de l'Espai (IEEC-CSIC),
Campus UAB, Facultat de Ciencies, Torre C5-Par-2a pl, E-08193 Bellaterra
(Barcelona), Spain\\
$^3$Dipartimento di Fisica, Universit\`a di Trento \\
and Istituto Nazionale di Fisica Nucleare \\
Gruppo Collegato di Trento, Italia
}


\begin{abstract}

We study all four types of finite-time future singularities emerging in 
late-time accelerating (effective quintessence/phantom) era from 
$\mathcal{F}(R,G)$-gravity, 
where $R$ and $G$ are the Ricci scalar and the Gauss-Bonnet invariant,
respectively.
As an explicit example of $\mathcal{F}(R,G)$-gravity, we
also investigate modified Gauss-Bonnet gravity, so-called $F(G)$-gravity.
In particular, we reconstruct the $F(G)$-gravity and
$\mathcal{F}(R,G)$-gravity models where accelerating cosmologies
realizing the finite-time future singularities emerge.
Furthermore, we discuss a possible way to cure the finite-time future
singularities in $F(G)$-gravity and $\mathcal{F}(R,G)$-gravity
by taking into account higher-order curvature corrections. 
The example of non-singular realistic modified Gauss-Bonnet gravity is 
presented. 
It turns out that adding such non-singular modified gravity to singular Dark 
Energy makes the combined theory to be non-singular one as well.

\end{abstract}


\maketitle

\section{Introduction}

Recent observations have implied that the current expansion of the universe
is accelerating~\cite{WMAP, SN1}.
There exist two broad categories to explain this phenomena~\cite{Peebles,
Copeland:2006wr, D-and-M, Nojiri:2006ri, rv-2, Sotiriou:2008rp, Lobo:2008sg,
Cap,C,Sami:2009jx}.
One is the introduction of ``dark energy'' in the framework of general
relativity. The other is the investigation of a modified gravitational
theory,
e.g., $f(R)$-gravity, in which the action is described by the Ricci scalar
$R$ plus an arbitrary function $f(R)$ of $R$ (for reviews,
see~\cite{Nojiri:2006ri, rv-2, Sotiriou:2008rp, Lobo:2008sg, Cap}).

It is known that accelerating Friedmann-Robertson-Walker (FRW) universe is
described by cosmological constant/quintessence/phantom Dark Energy.
In principle, Dark Energy (DE) could be described by scalar field theories,
fluid, modified gravity, etc. It is quite well-known that any of such DE
models may be represented as the effective fluid with corresponding
characteristics.
At the late-time accelerating stage of the FRW universe,
if the ratio of the effective pressure to the effective energy density
of the universe, i.e., the effective equation of state (EoS)
$w_\mathrm{eff}\equiv p_\mathrm{eff}/\rho_\mathrm{eff}$,
is larger than $-1$, it is the quintessence~\cite{Caldwell:1997ii,
Chiba:1997ej, Fujii:1982ms} (non-phantom)
phase. On the other hand, if $w_\mathrm{eff}$ is less than $-1$,
it is the phantom phase~\cite{Phantom phase} while effective cosmological
constant appears as DE when $w_\mathrm{eff}=-1$. Note that (non-transient)
phantom phase evolution usually ends up in Type I (Big Rip) future
singularity.
It is remarkable that many of the effective quintessence/phantom
DEs may bring the future universe evolution to finite-time
singularity.
The classification of such finite-time future singularities has been made
in Ref.~\cite{Nojiri:2005sx}. Some of these four types future singularities
are softer than other, for instance, not all characteristic quantities
(scale factor, effective pressure and energy-density) diverge in rip time.
There is not any qualitative difference between convenient DEs and
modified gravity in this respect.
For instance, the convenient parameter-dependent DE models may show all four
possible types of finite-time future singularity as demonstrated in
Refs.~\cite{Nojiri:2005sx, sing1, sing2}. On the same time it was
demonstrated in Refs.~\cite{Nojiri:2008fk, Odintsov} that $f(R)$-gravity
DE model may also bring the universe evolution to all four possible future
singularities (for first example of Big Rip (Type I) singularity in modified
gravity, see~\cite{Barrow:1990nv, Abdalla:2004sw, Briscese:2006xu}).
Furthermore, it is interesting that
$f(R)$ modified gravity may also provide the universal scenario to cure the
finite-time future singularity by adding, say,
$R^2$-term~\cite{Abdalla:2004sw, Nojiri:2008fk, Odintsov,
Capozziello:2009hc}
or non-singular viable $f(R)$-gravity~\cite{od}
(for related discussion of Type II
future singularity in particular $f(R)$-gravity and its curing by
$R^2$-term, see
Refs.~\cite{Kobayashi:2008tq,Dev:2008rx,Kobayashi:2008wc,Thongkool:2009js,
Babichev:2009td,Appleby:2009uf}).

It is clear that singular dark energy may lead to various instabilities in
the current universe cosmology, including black holes and stellar
astrophysics. In this respect it is very important
to list the singular dark energy models as well as try to indicate the
physical consequences of possible future singularity. Moreover, it is
desirable to construct the universal scenario to cure such singularities.
The interesting class of modified gravity models which may easily produce
the late-time acceleration epoch is string-inspired modified Gauss-Bonnet
gravity, so-called
$F(G)$-gravity~\cite{Nojiri:2005jg, F(G)-gravity, Zerbini,
Nojiri:2007bt},
where $F(G)$ is an arbitrary function of the Gauss-Bonnet invariant
$G=R^{2}-4R_{\mu\nu}R^{\mu\nu}+R_{\mu\nu\xi\sigma}R^{\mu\nu\xi\sigma}$
($R_{\mu\nu}$ and $R_{\mu\nu\xi\sigma}$ are the Ricci tensor and
the Riemann tensor, respectively). It is known that such class of models may
also lead to finite-time future singularity~\cite{Odintsov}.

In the present paper, as a generalized investigation of
Ref.~\cite{Odintsov},
we explore the $\mathcal{F}(R,G)$-gravity models
with realizing the finite-time future singularities
by using the reconstruction method of modified
gravity~\cite{Odintsov, Reconstruction},
where $\mathcal{F}(R,G)$ is an arbitrary function of $R$ and $G$.
The $\mathcal{F}(R,G)$-gravity
is a gravitational theory in a more general class of modified gravity and
includes $f(R)$-gravity and $F(G)$-gravity.
As an explicit example of $\mathcal{F}(R,G)$-gravity,
we also investigate $F(G)$-gravity and reconstruct the $F(G)$-gravity models
in which finite-time future singularities could appear in great detail.
It is shown that all four types of finite-time future singularity may
occur in such modified gravity.
In addition, we examine the possibility of the finite-time future
singularities in $F(G)$-gravity and $\mathcal{F}(R,G)$-gravity
being cured under higher-order curvature
corrections. The explicitly non-singular modified Gauss-Bonnet models is
proposed and it is shown that the finite-time future singularities may be
easily protected combining a singular theory with the non-singular one.
This suggests the universal scenario to cure the finite-time future
singularity in the same line as it was proposed in Ref.~\cite{od}.

The paper is organized as follows.
In Sec.~II,
we briefly review the model of $\mathcal{F}(R,G)$-gravity
and write down the gravitational field equations.
In addition, we classify the four types of the finite-time future
singularities.
In Sec.~III,
as a first step,
we investigate $F(G)$-gravity and reconstruct
the $F(G)$-gravity models where
finite-time future singularities may occur.
We also examine the finite-time future singularities in
realistic models of $F(G)$-gravity.
Next, in Sec.~IV we study the general
$\mathcal{F}(R,G)$-gravity models where
the finite-time future singularities occur.
Moreover, we explore the finite-time future singularities in
a realistic model of $\mathcal{F}(R,G)$-gravity.
In Sec.~V, we discuss a possible way to resolve the finite-time future
singularities in $F(G)$-gravity and $\mathcal{F}(R,G)$-gravity
by taking into account higher-order curvature corrections.
The non-singular theories are proposed.
It is shown that the addition of such non-singular effective dark energy
to the singular one may cure the singularity of the combined theory.
Hence, modified Gauss-Bonnet gravity may appear as the effective universal
regulator of finite-time future singularity not only for singular
alternative gravity but also for convenient singular DE. Finally,
conclusions
are given in Sec.\ VI.
The finite-time future singularities in a simple model of $f(R)$-gravity
are also examined in Appendix A. Furthermore,
a further argument on the asymptotic behavior of singular
models is presented in Appendix B.

We use units of $k_\mathrm{B} = c = \hbar = 1$ and denote the
gravitational constant $8 \pi G_{N}$ by
${\kappa}^2 \equiv 8\pi/{M_{\mathrm{Pl}}}^2$
with the Planck mass of $M_{\mathrm{Pl}} = G_{N}^{-1/2} =
1.2 \times 10^{19}$GeV.
A note on notation is that throughout the present paper,
$\alpha$, $\gamma$, $z$, $n$, $m$, $\delta$ and $\zeta$ are
constants unless we mention some conditions in regard to these
expressions.

\section{$\mathcal{F}(R,G)$-gravity}

In this section, we briefly review $\mathcal{F}(R,G)$-gravity
and derive the gravitational field equations.
Moreover, we classify the finite-time future
singularities into four types following ref.~\cite{Nojiri:2005sx}.

\subsection{The Model}

The action of $\mathcal{F}(R,G)$-gravity is given by
\begin{eqnarray}
S = \int d^4 x \sqrt{-g} \left[ \frac{\mathcal{F}(R,G)}{2\kappa^2}
+{\mathcal{L}}_{\mathrm{matter}} \right]\,,
\label{azione}
\end{eqnarray}
where $g$ is the determinant of the metric tensor $g_{\mu\nu}$
and ${\mathcal{L}}_{\mathrm{matter}}$ is the matter Lagrangian.

 From the action in Eq.~(\ref{azione}), the gravitational field equation is
derived as
\begin{eqnarray}
{\mathcal{F}}'_{R}\left( R_{\mu\nu}-\frac{1}{2}R g_{\mu\nu}\right)
\Eqn{=} \kappa^2 T^{(\mathrm{matter})}_{\mu \nu}
+\frac{1}{2}g_{\mu\nu} \left(\mathcal{F}-{\mathcal{F}}'_{R}R\right)
+{\nabla}_{\mu}{\nabla}_{\nu}
{\mathcal{F}}'_{R} -g_{\mu\nu} \Box {\mathcal{F}}'_{R}
\nonumber \\
&& \hspace{-30mm}
+\bigl(-2RR_{\mu\nu} +4R_{\mu\rho}R_{\nu}{}^{\rho}
-2R_{\mu}{}^{\rho\sigma\tau}R_{\nu\rho\sigma\tau}
+4g^{\alpha\rho}g^{\beta\sigma}R_{\mu\alpha\nu\beta}R_{\rho\sigma}
\bigr){\mathcal{F}}'_{G}
\nonumber \\
&& \hspace{-30mm}
+2\left({\nabla}_{\mu}{\nabla}_{\nu} {\mathcal{F}}'_{G} \right)R
-2g_{\mu \nu}\left(\Box {\mathcal{F}}'_{G} \right)R
+4\left(\Box {\mathcal{F}}'_{G} \right)R_{\mu \nu}
-4\left({\nabla}_{\rho}{\nabla}_{\mu} {\mathcal{F}}'_{G} \right)
R_{\nu}{}^{\rho}
\nonumber \\
&& \hspace{-30mm}
-4\left({\nabla}_{\rho}{\nabla}_{\nu} {\mathcal{F}}'_{G} \right)
R_{\mu}{}^{\rho}
+4g_{\mu \nu}\left({\nabla}_{\rho}{\nabla}_{\sigma}
{\mathcal{F}}'_{G} \right)R^{\rho\sigma}
-4\left({\nabla}_{\rho}{\nabla}_{\sigma} {\mathcal{F}}'_{G} \right)
g^{\alpha\rho}g^{\beta\sigma}R_{\mu\alpha\nu\beta}\,,
\label{eq:2.3}
\end{eqnarray}
where we have used the following expressions:
\begin{eqnarray}
{\mathcal{F}}'_{R} =
\frac{\partial \mathcal{F}(R,G)}{\partial R}\,, \quad
{\mathcal{F}}'_{G} =
\frac{\partial \mathcal{F}(R,G)}{\partial G}\,.
\label{eq:2.4}
\end{eqnarray}
Here, ${\nabla}_{\mu}$ is the covariant derivative
operator associated with $g_{\mu \nu}$,
$\Box \equiv g^{\mu \nu} {\nabla}_{\mu} {\nabla}_{\nu}$
is the covariant d'Alembertian for a scalar field, and
$T^{(\mathrm{matter})}_{\mu \nu} = \mathrm{diag} \left(\rho, p, p, p
\right)$
is the contribution to the energy-momentum tensor from
all ordinary matters
with $\rho$ and $p$ being the energy density and pressure of all ordinary
matters, respectively.

The most general flat FRW space-time is described by the metric
\begin{equation}
ds^{2}=-N^2(t)dt^{2}+a^{2}(t)d \mathbf{x}^{2}\,,
\label{metric}
\end{equation}
where $a(t)$ is the scale factor and $N(t)$ is an arbitrary function of $t$.
In what follows, we take $N(t)=1$.

In the FRW background, from $(\mu,\nu)=(0,0)$
and the trace part of $(\mu,\nu)=(i,j)$ $(i,j=1,\cdots,3)$ components in
Eq.~(\ref{eq:2.3}), we obtain the gravitational field equations:
\begin{equation}
\rho_{\mathrm{eff}}=\frac{3}{\kappa^{2}}H^{2}\,,
\quad
p_{\mathrm{eff}}=-\frac{1}{\kappa^{2}} \left( 2\dot H+3H^{2} \right)\,,
\label{GutenTag}
\end{equation}
where $\rho_{\mathrm{eff}}$ and $p_{\mathrm{eff}}$ are
the effective energy density and pressure of the universe, respectively, and
these are defined as
\begin{eqnarray}
\rho_{\mathrm{eff}} \Eqn{\equiv}
\frac{1}{{\mathcal{F}}'_{R}} \left\{ \rho +
\frac{1}{2\kappa^{2}}
\left[ \left( {\mathcal{F}}'_{R}R-\mathcal{F} \right)
-6H{\dot{\mathcal{F}}}'_{R}
+G{\mathcal{F}}'_{G}-24H^3{\dot{\mathcal{F}}}'_{G}
\right] \right\}\,,
\label{eq:2.7} \\
p_{\mathrm{eff}} \Eqn{\equiv}
\frac{1}{{\mathcal{F}}'_{R}} \biggl\{ p +
\frac{1}{2\kappa^{2}} \Bigl[
-\left( {\mathcal{F}}'_{R}R-\mathcal{F} \right)
+4H{\dot{\mathcal{F}}}'_{R}+2{\ddot{\mathcal{F}}}'_{R}
-G{\mathcal{F}}'_{G}
+16H\left(\dot{H} +H^2 \right){\dot{\mathcal{F}}}'_{G}
\nonumber \\
&& \hspace{10mm}
+8H^2 {\ddot{\mathcal{F}}}'_{G}
\Bigr]
\biggr\}\,.
\label{eq:2.8}
\end{eqnarray}
Here, $H=\dot{a}(t)/a(t)$ is the Hubble parameter and
the dot denotes the time derivative of $\partial/\partial t$.
For general relativity with $\mathcal{F}(R,G)=R$,
$\rho_{\mathrm{eff}} = \rho$ and $p_{\mathrm{eff}} = p$ and
therefore Eqs.~(\ref{eq:2.7}) and (\ref{eq:2.8}) are the FRW equations.
Consequently, Eqs.~(\ref{eq:2.7}) and (\ref{eq:2.8}) imply that
the contribution of modified gravity can formally be included in
the effective energy density and pressure of the universe.

\subsection{Four types of the finite-time future singularities}

We consider the case in which the Hubble parameter is expressed as
\begin{equation}
H=\frac{h}{(t_{0}-t)^{\beta}}+H_{0}\,,
\label{Hsingular}
\end{equation}
where $h$, $t_{0}$ and $H_{0}$ are positive constants,
$\beta$ is a constant, and $t<t_{0}$.
We can see that if
$\beta>0$, $H$ becomes singular in the limit $t\rightarrow t_{0}$.
Hence, $t_{0}$ is the time when a singularity appears.
On the other hand,
if $\beta<0$, even for non-integer values of $\beta$ some derivative of $H$
and therefore the curvature becomes singular~\cite{Odintsov}.
We assume $\beta\neq 0$ because $\beta=0$ corresponds to de Sitter space,
which has no singularity.

The finite-time future singularities can be classified
in the following way~\cite{Nojiri:2005sx}:
\begin{itemize}
\item Type I (Big Rip): for $t\rightarrow t_{0}$, $a(t)\rightarrow\infty$,
$\rho_\mathrm{{eff}}\rightarrow\infty$ and
$|p_\mathrm{{eff}}|\rightarrow\infty$. The case in which
$\rho_\mathrm{{eff}}$ and $p_\mathrm{{eff}}$ are finite at $t_{0}$ is also
included.
It corresponds to $\beta=1$ and $\beta>1$.
\item Type II (sudden~\cite{sudden}):
for $t\rightarrow t_{0}$, $a(t)\rightarrow a_{0}$,
$\rho_\mathrm{{eff}}\rightarrow\rho_{0}$ and $|p_\mathrm{{eff}}|
\rightarrow\infty$.
It corresponds to $-1<\beta<1$.
\item Type III: for $t\rightarrow t_{0}$, $a(t)\rightarrow a_{0}$,
$\rho_\mathrm{{eff}}\rightarrow\infty$ and
$|p_\mathrm{{eff}}|\rightarrow\infty$.
It corresponds to $0<\beta<1$.
\item Type IV: for $t\rightarrow t_{0}$, $a(t)\rightarrow a_{0}$,
$\rho_\mathrm{{eff}}\rightarrow 0$, $|p_\mathrm{{eff}}|
\rightarrow 0$
and higher derivatives of $H$ diverge.
The case in which $\rho$ and/or $p$ tend to finite values is
also included. It corresponds to
$\beta<-1$ but $\beta$ is not any integer number.
\end{itemize}
Here, $a_{0} (\neq 0)$ and $\rho_{0}$ are constants.
We note that in the present paper, we call singularities for $\beta=1$ and
those for $\beta>1$ as the ``Big Rip'' singularities and the ``Type I''
singularities, respectively.

\section{$F(G)$-gravity}

In this section, as an explicit example of $\mathcal{F}(R,G)$-gravity,
we first study $F(G)$-gravity~\cite{Zerbini, Nojiri:2005jg, F(G)-gravity,
Nojiri:2007bt}.
We reconstruct the $F(G)$-gravity models where
finite-time future singularities may occur.
In addition, we explore the finite-time future singularities in
realistic models of $F(G)$-gravity.

\subsection{The Model}

The action of $F(G)$-gravity is given by~\cite{Nojiri:2005jg}
\begin{equation}
S=\int d^{4}x \sqrt{-g} \left[ \frac{1}{2\kappa^2}
\left( R+F(G) \right)
+{\mathcal{L}}_{\mathrm{matter}}
\right]\,,
\label{one}
\end{equation}
which corresponds to the action in Eq.~(\ref{azione}) with
$\mathcal{F}(R,G) = R+F(G)$.

In the FRW background in Eq.~(\ref{metric}) with $N(t)=1$,
it follows from the action in Eq.~(\ref{one}) that
the equations of motion (EOM) for $F(G)$-gravity are given by~\cite{Zerbini}
\begin{eqnarray}
&& \hspace{-5mm}
24H^{3}\dot{F}'(G)+6H^{2}+F(G)-G F'(G)=2\kappa^{2}\rho\,,
\label{dancke} \\
&& \hspace{-5mm}
8H^{2}\ddot{F}'(G)+16H\dot{F}'(G)\left(\dot{H}+H^{2}\right)
+\left(4\dot{H}+6H^{2}\right)+F(G)-G F'(G)=-2\kappa^{2}p\,,
\label{bitte}
\end{eqnarray}
where the prime denotes differentiation with respect to $G$.
Moreover, we have
\begin{eqnarray}
R \Eqn{=} 6 \left(2H^{2}+\dot H \right)\,,
\label{eq:R} \\
G \Eqn{=} 24H^{2} \left( H^{2}+\dot H \right)\,.
\label{eq:G}
\end{eqnarray}
In this case,
$\rho_{\mathrm{eff}}$ and $p_{\mathrm{eff}}$ in the FRW
equations~(\ref{GutenTag}) take the form
\begin{eqnarray}
\rho_{\mathrm{eff}} \Eqn{=} \frac{1}{2\kappa^{2}} \left[ -F(G)+24H^{2}
\left(H^{2}+\dot H\right)F'(G)-24^{2}H^{4}
\left(2\dot H^{2}+H\ddot H+4H^{2}\dot H\right)F''(G)
\right]
\nonumber \\
&& \hspace{0mm}
+\rho\,,
\label{eq:rho-eff-1}
\\
p_{\mathrm{eff}} \Eqn{=} \frac{1}{2\kappa^{2}}\Bigl\{F(G)-24H^{2}
\left(H^{2}+\dot H\right)F'(G)
+(24)8H^{2}\Big[ 6\dot{H}^{3}+8 H \dot{H} \ddot{H}
+ 24\dot{H}^{2} H^2 + 6H^3\ddot{H}
\nonumber\\
& & + 8H^4\dot{H}+H^{2} \dddot{H} \Big]F''(G)+8(24)^{2}H^{4}
\left(2\dot{H}^2+H\ddot{H}+4H^{2}\dot{H}\right)^{2}F'''(G)\Bigr\}+p\,,
\label{eq:p-eff-1}
\end{eqnarray}
where we have used Eq.~(\ref{eq:G}).

We assume that the matter has a constant equation of state (EoS) parameter
$w \equiv p/\rho$. By combining the two equations in Eq.~(\ref{GutenTag}),
we obtain
\begin{equation}
\mathcal{G}(H,\dot{H}...)=
-\frac{1}{\kappa^2}\left[ 2\dot{H}+3(1+w)H^2 \right]\,,
\label{prime}
\end{equation}
where
\begin{equation}
\mathcal{G}(H,\dot{H}...)=p_{\mathrm{eff}}-w\rho_{\mathrm{eff}}\,.
\label{eq:def-mG}
\end{equation}
When a cosmology is given by $H=H(t)$, the right-hand side of
Eq.~(\ref{prime}) is described by a function of $t$.
If the function $\mathcal{G}(H,\dot{H}...)$ in Eq.~(\ref{eq:def-mG}),
which is the combination of $H$, $\dot{H}$, $\ddot{H}$ and
the higher derivatives of $H$,
reproduce the above function of $t$, this cosmology could be realized.
Hence, the function $\mathcal{G}(H,\dot{H}...)$ can be used to
judge whether the particular cosmology could be realized or
not~\cite{Odintsov}.
The form of $\mathcal{G}(H,\dot{H}...)$ is determined by the
gravitational theory which one considers.
In the case of $F(G)$-gravity,
by substituting Eqs.~(\ref{eq:rho-eff-1}) and (\ref{eq:p-eff-1}) into
Eq.~(\ref{eq:def-mG}), we find
\begin{eqnarray}
\mathcal{G}(H,\dot{H}...)
\Eqn{=}
\frac{1}{2\kappa^{2}} \biggl\{ (1+w)F(G)-24(1+w)H^2\left(H^2+\dot{H}\right)
F'(G)+8(24)H^{2} \biggl[ 6\dot{H}^{3}
\nonumber \\
&&
+8 H \dot{H}\ddot{H}
+ 6(4+w)\dot{H}^{2} H^2 + 3(2+w)H^3 \ddot{H}+4(2+3w)H^4 \dot{H}+H^{2}
\dddot{H} \biggr] F''(G)
\nonumber\\
& & + 8(24)^{2}H^{4}
\left(2\dot{H}^2+H\ddot{H}+4H^{2}\dot{H}\right)^{2}F'''(G) \biggr\}\,.
\label{MuyBien}
\end{eqnarray}

\subsection{Finite-time future singularities in $F(G)$-gravity}

We investigate the $F(G)$-gravity models in which the
finite-time future singularities could occur, when the form of $H$
is taken as Eq.~(\ref{Hsingular}).
To find such $F(G)$-gravity models, we use
the reconstruction method of modified gravity~\cite{Odintsov,
Reconstruction}.
By using proper functions $P(t)$ and $Q(t)$ of a scalar field $t$
which we identify with the cosmic time,
the action in Eq.~({\ref{one}}) can be rewritten to
\begin{equation}
S=\int d^{4}x \sqrt{-g} \left[
\frac{1}{2\kappa^2}\left( R+P(t)G+Q(t) \right)
+{\mathcal{L}}_{\mathrm{matter}}\right]\,.
\label{onebis}
\end{equation}
The variation with respect to $t$ yields
\begin{equation}
\frac{d P(t)}{dt}G+\frac{d Q(t)}{dt}=0\,,
\label{PQ}
\end{equation}
from which we can find $t=t(G)$.
By substituting $t=t(G)$ into Eq.~(\ref{onebis}), we
find the action in terms of $F(G)$
\begin{equation}
F(G)=P(t)G+Q(t)\,.
\label{F}
\end{equation}

We describe the scale factor as
\begin{eqnarray}
a(t)= \bar{a} \exp \left( g(t) \right)\,,
\label{eq:Scale-Factor}
\end{eqnarray}
where $\bar{a}$ is a constant and $g(t)$ is a proper function.
By using Eqs.~(\ref{dancke}), Eq.~(\ref{bitte}), (\ref{eq:Scale-Factor}),
the
matter conservation law $\dot \rho+3 H (\rho+p)=0 $ and then neglecting the
contribution from matter,
we get the differential equation
\begin{equation}
2 \frac{d}{d t} \left( \dot{g}^2(t)  \frac{d P(t)}{d t} \right) -2
\dot{g}^{3}(t) \frac{d P(t)}{d t} + \ddot{g}(t)=0\,.
\label{P}
\end{equation}
By using the first EOM for $F(G)$-gravity in Eq.~(\ref{dancke}),
$Q(t)$ is given by
\begin{equation}
Q(t)= -24 \dot{g}^{3}(t)\frac{d P(t)}{d t}-6\dot{g}^2(t)\,.
\label{Q}
\end{equation}

\subsubsection{Big Rip singularity}

First, we examine the Big Rip singularity.
If $\beta=1$ in Eq.~(\ref{Hsingular}) with $H_{0}=0$,
$H$ and $G$ are given by
\begin{eqnarray}
H \Eqn{=} \frac{h}{(t_{0}-t)}\,,
\label{H} \\
G \Eqn{=} \frac{24h^3}{(t_{0}-t)^4}(1+h)\,.
\label{G}
\end{eqnarray}
The most general solution of Eq.~(\ref{P}) is given by
\begin{equation}
P(t)=\frac{1}{4h(h-1)}(2t_{0}-t)t+c_{1}\frac{(t_{0}-t)^{3-h}}{3-h}+c_{2}\,,
\end{equation}
where $c_{1}$ and $c_{2}$ are constants. From Eq.~($\ref{Q}$),
we get
\begin{equation}
Q(t)=-\frac{6h^{2}}{(t_{0}-t)^2}-\dfrac{24 h^{3} \left[
\frac{(t_{0}-t)}{2h(h-1)}-c_{1}(t_{0}-t)^{2-h} \right]}{(t_{0}-t)^3}\,.
\end{equation}
Furthermore, from Eq.~(\ref{PQ}) we obtain
\begin{equation}
t= \left[ \dfrac{24(h^{3}+h^{4})}{G} \right]^{1/4}+t_{0}\,,
\end{equation}
which is consistent with Eq.~(\ref{G}). By solving Eq.~(\ref{F}), we find
the most general form of $F(G)$ which realizes the Big Rip singularity
\begin{equation}
F(G)=\frac{\sqrt{6h^{3}(1+h)}}{h(1-h)}\sqrt{G}
+c_{1}G^{\frac{h+1}{4}}+c_{2}G\,.
\label{Garr}
\end{equation}
This is an exact solution of Eq.~(\ref{prime}) in the case of Eq.~(\ref{H}).
In general, if for large values of $G$,
$F(G)\sim \alpha G^{1/2}$, where
$\alpha (\neq 0)$ is a constant,
the Big Rip singularity could appear for any value of $h\neq 1$.
In the case of $h=1$, the solution of $\mathcal{G}(H, \dot{H}...)$ is zero
for
$F(G)=\alpha G^{1/2}$. Note that $\alpha G^{(1+h)/4}$ is an invariant with
respect to the Big Rip solution.

In the case of $h=1$, it is possible to find
another exact solution for $P(t)$
\begin{equation}
P(t)=\alpha(t_{0}-t)^{q}\ln \left[ \gamma(t_{0}-t)^{z} \right]\,,
\end{equation}
where $\gamma ( > 0 )$ is a positive constant and $q$ and $z$ are
constants.
The equation (\ref{P}) is satisfied for the case of Eq.~(\ref{H})
if $q=3-h=2$
(and therefore $h=1$) and $z\alpha=-1/4$. From Eq.~(\ref{Q}), we have
\begin{equation}
Q(t)=-\frac{12}{(t_{0}-t)^{2}}\ln \left[ \gamma (t_{0}-t) \right]\,.
\label{schiappa}
\end{equation}
The form of $F(G)$ is given by
\begin{equation}
F(G)= \frac{\sqrt{3}}{2}\sqrt{G} \ln(\gamma G)\,.
\end{equation}
This is another exact solution of Eq.~(\ref{prime}) for $H=1/(t_{0}-t)$.
In general,
if for large values of $G$,
$F(G)\sim \alpha\sqrt{G} \ln(\gamma G)$
with $\alpha>0$ and $\gamma>0$,
the Big Rip singularity could appear.
The same result is found for
$F(G)\sim \alpha\sqrt{G} \ln(\gamma G^{z}+G_{0})$
with $\alpha>0$, $\gamma>0$ and $z>0$, where $G_{0}$ is a constant.

\subsubsection{Other types of singularities}

Next, we investigate the other types of singularities.
If $\beta\neq 1$, Eq.~(\ref{Hsingular}) with $H_{0}=0$
implies that the scale factor $a(t)$ behaves as
\begin{equation}
a(t)
= \exp \left[ \frac{h(t_{0}-t)^{1-\beta}}{\beta-1} \right]\,.
\end{equation}

We consider the case in which $H$ and $G$ are given by
\begin{eqnarray}
H \Eqn{=} \frac{h}{(t_{0}-t)^{\beta}}\,,
\quad
\beta>1\,,
\label{HII} \\
G \Eqn{\sim} \frac{24h^4}{(t_{0}-t)^{4\beta}}\,.
\end{eqnarray}

A solution of Eq.~(\ref{P}) in the limit $t\rightarrow t_{0}$ is given by
\begin{equation}
P(t)\simeq \frac{\alpha}{(t_{0}-t)^{z}}
\end{equation}
with $z=-2\beta$ and $\alpha=-1/4h^{2}$. The form of $F(G)$ is expressed as
\begin{equation}
F(G)=-12\sqrt{\frac{G}{24}}\,.
\label{primo}
\end{equation}
Hence, if for large values of $G$, $F(G)\sim -\alpha\sqrt{G}$
with $\alpha>0$, a Type I singularity could appear.

When $\beta<1$, the forms of $H$ and $G$ are given by
\begin{eqnarray}
H \Eqn{=} \frac{h}{(t_{0}-t)^{\beta}}\,,
\quad
0<\beta<1\,,
\label{Hsuper} \\
G \Eqn{\sim} \frac{24h^3\beta}{(t_{0}-t)^{3\beta+1}}\,.
\end{eqnarray}
An asymptotic solution of Eq.~(\ref{P}) in the limit $t\rightarrow t_{0}$ is
given by
\begin{equation}
P(t)\simeq \frac{\alpha}{(t_{0}-t)^{z}}
\end{equation}
with $z=-(1+\beta)$ and $\alpha=1/2h(1+\beta)$. The form of $F(G)$
becomes
\begin{equation}
F(G)=\frac{6h^{2}}{(\beta+1)}(3\beta+1)\left(\frac{|G|}{24h^{3}|\beta|}
\right)^{2\beta/(3\beta+1)}\,.
\label{secondo}
\end{equation}
Hence, if for large values of $G$, $F(G)$ has the form
\begin{equation}
F(G)\sim \alpha |G|^{\gamma}\,,
\quad
\gamma = \frac{2\beta}{3\beta+1}\,,
\label{trentatre}
\end{equation}
with $\alpha>0$ and $0<\gamma <1/2$,
we find $0<\beta<1$ and a Type III singularity could emerge.

If for $G\rightarrow-\infty$,
$F(G)$ has the form in Eq.~(\ref{trentatre})
with $\alpha>0$ and $-\infty<\gamma<0$,
we find $-1/3<\beta<0$ and
a Type II (sudden) singularity could appear.
Moreover,
if for $G\rightarrow 0^{-}$,
$F(G)$ has the form in Eq.~(\ref{trentatre})
with $\alpha<0$ and $1<\gamma<\infty$,
we obtain $-1<\beta<-1/3$ and a Type II singularity could occur.

If for $G\rightarrow 0^{-}$, $F(G)$ has the form in Eq.~(\ref{trentatre})
with $\alpha>0$ and $2/3<\gamma<1$,
we obtain $-\infty<\beta<-1$ and a Type IV singularity could appear.
We also require that $\gamma\neq2n/(3n-1)$,
where $n$ is a natural number.

We can generate all the possible Type II singularities
as shown above except in the case $\beta=-1/3$,
i.e., $H=h/(t_{0}-t)^{1/3}$.
In this case, we have the following form of $G$:
\begin{equation}
G=24h^{3}\beta+24h^{4}(t_{0}-t)^{4/3}<0\,.
\end{equation}
To find $t$ in terms of $G$, we must
consider the whole expression of $G$ by taking into account also the low
term
of $(t_{0}-t)$. We obtain
\begin{equation}
F(G)\simeq \frac{1}{4 \sqrt{6}h^{3}}G(G+8h^{3})^{1/2}
+\frac{2}{\sqrt{6}}(G+8h^{3})^{1/2}\,,
\end{equation}
which satisfies Eq.~(\ref{prime}) in the limit $t\rightarrow t_{0}$.
As a consequence,
the specific model $F(G)= \sigma_1 G(G+c_{3})^{1/2}+ \sigma_2
(G+c_{3})^{1/2}$,
where $\sigma_1$, $\sigma_2$ and $c_{3}$
are positive constants, can generate a Type II singularity.

\subsection{Realistic models of $F(G)$-gravity}

Here, we study the realistic models of $F(G)$-gravity,
which reproduce the current acceleration, namely~\cite{Nojiri:2007bt}
\begin{eqnarray}
F_{1}(G) \Eqn{=} \frac{a_{1}G^{n}+b_{1}}{a_{2}G^{n}+b_{2}}\,,
\label{uno} \\
F_{2}(G) \Eqn{=} \frac{a_{1}G^{n+N}+b_{1}}{a_{2}G^{n}+b_{2}}\,,
\label{due} \\
F_{3}(G) \Eqn{=} a_{3} G^{n}(1+b_{3} G^{m})\,,
\label{terzo}
\end{eqnarray}
where $a_{1}$, $a_{2}$, $b_{1}$, $b_{2}$, $a_{3}$, $b_{3}$, $n$, $N$ and $m$
are constants.
In the following, we always assume $n > 0$. For the model~(\ref{due}),
Types I, II and III singularities may be present. In fact, for  $N=1/2$, one
could have Big Rip singularities,
since in this case, in the limit large $G$, Eq. ~(\ref{due}) gives
$\alpha G^{1/2}$. Thus, as discussed
in Subsection III. B, one has a Big Rip singularity. Moreover, again with
$N=1/2$, if $a_1/a_2 <0$, Eq.~(\ref{due})
for large value of $G$, leads to $-\alpha G^{1/2}$ with $ \alpha >0$ and
thus
Type I singularity could appear.
If $n$ and $N$ are integers and $n+N >0$, for large and negative value of
$G$, $F_2(G)\sim a_1/a_2 G^N$.
As a result, a Type II singularity could
appear, when $-n <N<0$, $N$ even and $a_1/a_2>0$ or $N$ odd and   $a_1/a_2
<0$ (see Eq.~(\ref{trentatre}) and the related discussion). If $0<N< 1/2$
and $a_{1}/a_{2}>0$, we have the Type III singularity
(see Eq.~(\ref{trentatre})). When $G\rightarrow 0^{-}$, we do not recover
any example of singularity of
the preceding subsection.

If there exists any singularity solution,
it must be consistent with Eq.~(\ref{prime}).
The behavior of Eq.~(\ref{MuyBien}) takes two asymptotic forms
which depend on the parameter of $\beta$ as follows:
\begin{itemize}
\item
Case of $\beta \geq 1$:
In the limit $t\rightarrow t_{0}$,
we find
\begin{equation}
\mathcal{G}(H, \dot{H}...)\sim
\alpha
F(G)+\frac{\gamma}{(t_{0}-t)^{4\beta}}F'(G)+\frac{\delta}{(t_{0}-t)^{7\beta+
1}}F''(G)+\frac{\zeta}{(t_{0}-t)^{10\beta+2}}F'''(G)
\,,
\label{MeinGott}
\end{equation}
where $\delta$ and $\zeta$ are constants.
To realize a singularity, from Eq.~(\ref{prime}) we must have
\begin{equation}
\mathcal{G}(H, \dot{H}...)\sim
-\frac{3(1+w)h^{2}}{\kappa^{2}(t_{0}-t)^{2\beta}}\,.
\label{Luftwaffe}
\end{equation}
Hence,
if for $G\sim 24h^{4}/(t_{0}-t)^{4\beta}$ with $\beta \geq 1$,
the highest term of Eq.~($\ref{MeinGott}$) is proportional to
$1/(t_{0}-t)^{2\beta}$, it is possible to have a Type I singularity.
This condition is necessary and not sufficient.
Another very important condition that must be satisfied is the concordance
of the signs in Eq.~(\ref{Luftwaffe}),
which depends on the parameters of the model.
\item
Case of $\beta<1$:
In the limit $t\rightarrow t_{0}$
we obtain
\begin{equation}
\mathcal{G}(H, \dot{H}...)\sim
\alpha
F(G)+\frac{\gamma}{(t_{0}-t)^{3\beta+1}}F'(G)+
\frac{\delta}{(t_{0}-t)^{5\beta+3}}F''(G)+\frac{\zeta}{(t_{0}-t)^{8\beta+4}}
F'''(G)
\,.
\label{ja}
\end{equation}
To realize a singularity, from Eq.~(\ref{prime}) we must have
\begin{equation}
\mathcal{G}(H, \dot{H}...)\sim
-\frac{2\beta h}{\kappa^{2}(t_{0}-t)^{\beta+1}}\,.
\label{nein}
\end{equation}
Thus, if for $G\sim 24h^{3}\beta/(t_{0}-t)^{3\beta+1}$ with $\beta<1$, the
highest term of Eq.~($\ref{ja}$) is proportional to $1/(t_{0}-t)^{\beta+1}$,
it is possible to have a Type II, III or IV singularity. Also this condition
is necessary and not sufficient.
\end{itemize}

We see that the model in Eq.~(\ref{due}) with
$n>0$ and $N>0$
is not able to
realize a Type IV singularity
because for $\beta<-1$
the right-hand side of Eq.~(\ref{nein})
tends to zero and the left-hand side of Eq.~(\ref{nein})
tends to a constant ($F_{2}(G)\sim b_{1}/b_{2}$). Nevertheless,
it is possible to have a Type II singularity for $0<\beta<-1/3$.
If $n>0$ and $N>0$, we get
\begin{eqnarray}
&&
F_{2}(G)\sim\frac{b_{1}}{b_{2}}\,,
\quad
F_{2}'(G)\sim-n\frac{b_{1}a_{2}}{b_{2}^{2}}G^{n-1}\,,
\quad
F_{2}''(G)\sim-n\frac{b_{1}a_{2}}{b_{2}^{2}}(n-1)G^{n-2}\,,
\nonumber \\
&& \hspace{0mm}
F_{2}'''(G)\sim-n\frac{b_{1}a_{2}}{b_{2}^{2}}(n-1)(n-2)G^{n-3}\,.
\end{eqnarray}
It can be shown that, under the requirement $n>1$ (the relation between $n$
and $\beta$ is $n=2\beta/(3\beta+1)$), the asymptotic behavior of
Eq.~(\ref{ja}) when $G\simeq 24 h^{3}\beta/(t_{0}-t)^{3\beta+1}$ is
proportional to $1/(t_{0}-t)^{\beta+1}$ and
therefore it is possible to realize the Type II singularity.
\begin{itemize}
\item
For $N=1$ and $n=2$, $\mathcal{G}(H,\dot H...)\sim
(24 h^{5})b_{1}a_{2}/b_{2}^{2}$ when $\beta=-1/2$.
Hence,
if $b_{1}a_{2}>0$, the model can become singular when $G\rightarrow 0^{-}$
(Type II singularity).
\item
For $N=1$ and $n=3$, $\mathcal{G}(H,\dot H...)\sim
-b_{1}a_{2}/b_{2}^{2}$ when $\beta=-3/7$.
Thus, if $b_{1}a_{2}<0$, the model can become singular when
$G\rightarrow 0^{-}$ (Type II singularity).
\end{itemize}

In a certain sense, the model $F_{1}(G)$ in Eq.~(\ref{uno}) is a particular
case of Eq.~(\ref{due}). For large values of $G$, it tends to a constant
with
velocity being zero,
so that it is impossible to find singularities (it is well known that
$R+constant$ is free of singularities, according to the
$\Lambda$CDM model).
Nevertheless, similarly to the above,
a Type II singularity can occur when $G\rightarrow 0^{-}$
for $n>1$.
For example,
if $n=2$, $\mathcal{G}(H,\dot H)\sim
(24 h^{5}/b_{2}^{2})(b_{1}a_{2}-a_{1}b_{2})$ for
$\beta=-1/2$. If $b_{1}a_{2}-a_{1}b_{2}>0$, the model can become singular
when $G\rightarrow 0^{-}$.

With regard to $F_{3}(G)$ in Eq.~(\ref{terzo}), it is interesting to find
the conditions on $m$, $n$, $a_{3}$ and $b_{3}$ for
which we
do not have any type of singularities. When $G\rightarrow \pm\infty$ or
$G\rightarrow 0^{-}$, it is possible to write this model in the form
$F(G)\sim\alpha G^{\gamma}$, which we have investigated on in the
preceding subsection.
We do not consider the trivial case $n=m$.
The no-singularity conditions follow directly from the results of the
preceding subsection as complementary  conditions to the singularity ones:
\begin{itemize}
\item
Case (A):
$n>0$, $m>0$, $n\neq 1$ and $m\neq 1$.
We avoid any singularity if $0<n+m<1/2$ and $a_{3}b_{3}<0$; $n+m>1/2$, $n>1$
and $a_{3}>0$; $n+m>1/2$, $2/3<n<1$ and $a_{3}<0$; $n+m>1/2$,
$0<n \leq 2/3$ and if $n=1/2$, $a_{3}>0$.
\item
Case (B):
$n>0$, $m<0$ and $n\neq 1$.
We avoid any singularity if $0<n<1/2$ and
$a_{3}<0$; $n>1/2$, $n+m> 1$ and $a_{3}b_{3}>0$; $n>1/2$, $2/3<n+m < 1$
and $a_{3}b_{3}<0$; $n>1/2$, $n+m \leq 2/3$ and if $n+m=1/2$,
$a_{3}b_{3}>0$.
\item
Case (C):
$n<0$, $m>0$ and $m\neq 1$.
We avoid any singularity if $m+n>1/2$; $m+n<1/2$ and
$a_{3}b_{3}<0$.
\item
Case (D):
$n<0$ and $m<0$.
We avoid any singularity if $a_{3}<0$.
\end{itemize}

We end this subsection considering the following realistic model, again for
$n>0$,
\begin{eqnarray}
F_{4}(G) = G^\alpha \frac{a_{1}G^{n}+b_{1}}{a_{2}G^{n}+b_{2}}\,.
\label{QUARTO}
\end{eqnarray}
Since for large $G$, one has $F_{4}(G) \simeq a_1/a_2G^\alpha $ and for
small $G$, one has
$F_{4}(G) \simeq b_1/b_2G^\alpha $, the analysis of Subsection III. B leads
to the absence of any type of singularities
for
\begin{equation}
\frac{1}{2} < \alpha < \frac{2}{3}\,.
\label{sf}
\end{equation}
In fact, for this range of values, the asymptotic behavior of the right-hand
side of Eq.~(\ref{prime}) is different
from the asymptotic behavior of its left-hand side on the singularity
solutions. Thus, Eq.~(\ref{QUARTO}) provides an example of realistic model
free of all possible singularities when Eq.~(\ref{sf}) is satisfied,
independently of the coefficients. Moreover, this model suggests the
universal scenario to cure finite-time future singularity.
Adding above model to any singular Dark Energy (in the same way as adding
$R^2$-term~\cite{Abdalla:2004sw,Odintsov,Nojiri:2008fk}) results in combined
non-singular model. Hence, unlike to convenient DE which may be singular or
not, (non-singular) modified gravity may suggest the universal recipe to
cure the finite-time future singularity. In this respect, modified gravity
seems to be more fundamental theory than convenient DEs.


\section{Finite-time future singularities in $\mathcal{F}(R,G)$-gravity}

In this section, we consider the finite-time future singularities
in $\mathcal{F}(R,G)$-gravity.
We reconstruct the $\mathcal{F}(R,G)$-gravity models with producing
the finite-time future singularities.
Furthermore, we examine the finite-time future singularities in
a realistic model of $\mathcal{F}(R,G)$-gravity.

\subsection{Formalism}

We study the pure gravitational action of $\mathcal{F}(R,G)$-gravity,
i.e., the action in Eq.~(\ref{azione}) without
${\mathcal{L}}_{\mathrm{matter}}$.
In this case, it follows from Eqs.~(\ref{eq:2.7}) and (\ref{eq:2.8}) that
the EOM of $F(R,G)$-gravity are given by~\cite{Zerbini}
\begin{eqnarray}
&&
24H^{3}{\dot{\mathcal{F}}}'_{G}
+6H^{2}{\mathcal{F}}'_{R}
+6H{\dot{\mathcal{F}}}'_{R}+(\mathcal{F}-R{\mathcal{F}}'_{R}
-G{\mathcal{F}}'_{G})=0\,,
\label{un} \\
&&
8H^{2}{\ddot{\mathcal{F}}}'_{G}+2{\ddot{\mathcal{F}}}'_{R}
+4H{\dot{\mathcal{F}}}'_{R}+16H{\dot{\mathcal{F}}}'_{G}(\dot H+H^{2})
\nonumber \\
&& \hspace{14mm}
+{\mathcal{F}}'_{R}(4\dot H+6H^{2})
+\mathcal{F}-R {\mathcal{F}}'_{R}
-G {\mathcal{F}}'_{G}=0\,.
\label{dos}
\end{eqnarray}
In the case of pure gravity,
these two equations are linearly dependent.

Now, similarly to the previous section,
by using proper functions $P(t)$, $Z(t)$ and $Q(t)$ of a
scalar field which is identified with the time $t$,
we can rewrite the action in Eq.~(\ref{azione}) without
${\mathcal{L}}_{\mathrm{matter}}$ to
\begin{equation}
S=\frac{1}{2\kappa^2}\int d^{4}x \sqrt{-g} \left( P(t)R+Z(t)G+Q(t)
\right)\,.
\label{azionemodificata}
\end{equation}
By the variation with respect to $t$, we obtain
\begin{equation}
P'(t)R+Z'(t)G+Q'(t)=0\,,
\label{t}
\end{equation}
from which in principle it is possible to find $t=t(R,G)$.
Here, the prime denotes differentiation with respect to $t$.
By substituting $t=t(R,G)$ into Eq.~(\ref{azionemodificata}),
we find the action in terms of $\mathcal{F}(R,G)$
\begin{equation}
\mathcal{F}(R,G)=P(t)R+Z(t)G+Q(t)\label{F(R,G)}\,.
\end{equation}
By using the conservation law and Eq.~(\ref{un}), we get the differential
equation
\begin{equation}
P''(t)+4 \dot{g}^{2}(t)Z''(t)-\dot g(t)P'(t)+(8 \dot{g} \ddot{g}
-4 \dot{g}^{3}(t))Z'(t)+2 \ddot{g}(t)P(t)=0\,,
\label{Pi}
\end{equation}
where we have used the expression of
the scale factor in Eq.~(\ref{eq:Scale-Factor})
and the Hubble parameter $H(t)=\dot{g}(t)$.
By using Eq.~(\ref{un}), $Q(t)$ becomes
\begin{equation}
Q(t)=-24\dot{g}^{3}(t)Z'(t)-6\dot{g}^{2}(t)P(t)-6\dot{g}(t)P'(t)\,.
\label{Qu}
\end{equation}
In general, if $P(t)\neq 0$, $\mathcal{F}(R,G)$ can be written in the
following form:
\begin{equation}
 \mathcal{F}(R,G)=R g(R,G)+f(R,G)\,,
\label{zuzzurellone}
\end{equation}
where $g(R,G)\neq 0$ and $f(R,G)$ are generic functions of $R$ and $G$. From
Eqs.~(\ref{un}) and (\ref{dos}), we obtain
\begin{eqnarray}
\rho_{\mathrm{eff}} \Eqn{=} -\frac{1}{2\kappa^{2}g(R,G)}\biggl[
24H^{3}{\dot{\mathcal{F}}}'_{G}
+6H^{2}\left( R\frac{d g(R,G)}{d R}+\frac{d f(R,G)}{d R} \right)
+6H{\dot{\mathcal{F}}}'_{R}
\nonumber \\
&& \hspace{25mm}
+(\mathcal{F}-R{\mathcal{F}}'_{R}
-G{\mathcal{F}}'_{G})\biggr]
\label{tic}
\end{eqnarray}
and
\begin{eqnarray}
p_{\mathrm{eff}} \Eqn{=} \frac{1}{2\kappa^{2}g(R,G)}\biggl[
8H^{2}{\ddot{\mathcal{F}}}'_{G}+2{\ddot{\mathcal{F}}}'_{R}
+4H{\dot{\mathcal{F}}}'_{R}+16H{\dot{\mathcal{F}}}'_{G}(\dot H+H^{2})
\nonumber \\
& & +\left(R\frac{d g(R,G)}{d R}+\frac{d f(R,G)}{d R}\right)(4\dot H+6H^{2})
+
\mathcal{F}-R {\mathcal{F}}'_{R}-G {\mathcal{F}}'_{G}\biggr]\,,
\label{tac}
\end{eqnarray}
respectively,
where $\rho_{\mathrm{eff}}$ and $p_{\mathrm{eff}}$ are given by the
expressions in (\ref{GutenTag}).
As a consequence,
we recover the same formalism of Sec.~III as
\begin{eqnarray}
\hspace{-5mm}
\mathcal{G}(H,\dot{H}...) \Eqn{=}
p_{\mathrm{eff}}-w\rho_{\mathrm{eff}}
\nonumber \\
\Eqn{=}
\frac{1}{2\kappa^{2}g(R,G)} \biggl\{ (1+w)(\mathcal{F}
-R{\mathcal{F}}'_{R}-G{\mathcal{F}}'_{G})
\nonumber \\
\hspace{-5mm}
&&
+\left(R\frac{dg(R,G)}{dR}+\frac{df(R,G)}{dR}\right)
\left[6H^2(1+w)+4\dot{H}\right]
\nonumber \\
\hspace{-5mm}
& &
+ H{\dot{\mathcal{F}}}'_{R}(4+6w)
+8H{\dot{\mathcal{F}}}'_{G}\left[2\dot{H}+ H^{2}(2+3w)\right]
+2{\ddot{\mathcal{F}}}'_{R}+8 H^{2}{\ddot{\mathcal{F}}}'_{G} \biggr\}\,,
\label{Feldwebel}
\end{eqnarray}
where $w$ is the constant EoS parameter of matter. The use of this
equation requires that $g(R,G)\neq 0$ on the solution.
The equation for $\mathcal{G}(H,\dot{H}...)$ is given by
Eq.~(\ref{prime}).

\subsection{Finite-time future singularities}

We examine the $\mathcal{F}(R,G)$-gravity models in which
the finite-time future singularities could appear.

\subsubsection{Big Rip singularity}

First, we investigate the Big Rip singularity.
If $\beta=1$ in Eq.~(\ref{Hsingular}) with $H_{0}=0$,
we have
\begin{eqnarray}
H \Eqn{=} \frac{h}{t_{0}-t}\,,
\label{BIGRIP} \\
R \Eqn{=} \frac{6 h}{(t_{0}-t)^{2}}(2h+1)\,, \\
G \Eqn{=} \frac{24 h^{3}}{(t_{0}-t)^{4}}(1+h)\,,
\end{eqnarray}
with $h>0$. A simple (trivial) solution of Eq.~(\ref{Pi}) is given by
\begin{eqnarray}
P(t) \Eqn{=} \alpha(t_{0}-t)^{z}\,, \\
Z(t) \Eqn{=} \delta(t_{0}-t)^{x}\,,
\end{eqnarray}
with $\alpha$ and $\delta$ being constants, where $x=3-h$ and $z$ is given
by
\begin{equation}
z_{\pm}
=\dfrac{1-h\pm \sqrt{h^{2}-10 h+1}}{2}\,.
\end{equation}
Thus, the most general solution of $P(t)$ is expressed as
\begin{equation}
P(t)=\alpha_{1}(t_{0}-t)^{z_{+}}+\alpha_{2}(t_{0}-t)^{z_{-}}\,,
\end{equation}
where $\alpha_{1}$ and $\alpha_{2}$ are constants. From Eq.~(\ref{Qu}),
we have
\begin{equation}
Q(t)=\dfrac{24h^{3}\delta(3-h)}{(t_{0}-t)^{h+1}}+\dfrac{6h\alpha_{1}(z_{+}-h
)}{(t_{0}-t)^{2-z_{+}}}+\dfrac{6h\alpha_{2}(z_{-}-h)}{(t_{0}-t)^{2-z_{-}}}\,
.
\end{equation}
Under the condition $0<h<5-2\sqrt{6}$ or $h>2+\sqrt{6}$, the solution of
$\mathcal{F}(R,G)$ (by absorbing some factor into the constants) is given by
\begin{equation}
\mathcal{F}(R,G)=\alpha_{1}R^{1-z_{+}/2}+\alpha_{2}R^{1-z_{-}/2}
+\delta G^{\frac{h+1}{4}}
\,.
\label{trivial}
\end{equation}
If $\delta=0$, we find a well-known result of $f(R)$-gravity.
$G^{\frac{h+1}{4}}$ is an invariant of the Big Rip solution
in a $F(G)$-gravity theory
and it is a solution in a general $\mathcal{F}(R,G)$-gravity theory.
Note that
$1-z_{\pm}\neq 1$.

Another exact solution of Eq.~(\ref{Pi}) is given by
\begin{eqnarray}
P(t) \Eqn{=} \frac{\alpha}{(t_{0}-t)^{z}}\,, \\
Z(t) \Eqn{=} \frac{\delta}{(t_{0}-t)^{x}}\,,
\end{eqnarray}
where $\delta$ and $x$ are constants, $z=x+2$ and $\alpha$
is given by
\begin{equation}
\alpha=\dfrac{4 h^{2}\delta x (h-x-3)}{x^{2}+(5-h)x+6}\,.
\end{equation}
 From Eq.~(\ref{Qu}), we find
\begin{equation}
Q(t)=-\frac{6h}{(t_{0}-t)^{x+4}}\left[4h^{2}\delta x+\alpha(x+2+h)
\right]\,.
\end{equation}
The solution of Eq.~(\ref{t}) is given by
\begin{eqnarray}
&& \hspace{-15mm}
t_{0}-t
= f(R,G)
\nonumber \\
&& \hspace{-15mm}
=\left\{\dfrac{-\alpha(x+2)R\pm
\sqrt{\alpha^{2}(x+2)^{2}R^{2}+24h\left[4h^{2}\delta x+\alpha(x+2+h)\right]
(x+4)\delta x G}}{2\delta x G}\right\}^{1/2}\,,
\label{radice}
\end{eqnarray}
with $x\neq 0$ and $\delta\neq 0$.

To have real solutions, we must require that the arguments of the
roots in Eq.~(\ref{radice}) are positive.
For $h>0$, the principal cases are as follows:
\begin{itemize}
\item
Case (1):
$x>0$, $\delta>0$, $1+x \leq h < x+5+\frac{6}{x}$.
We must use the sign $+$ in (\ref{radice}).
\item
Case (2):
$-\frac{3}{2} \leq x<0$, $\delta<0$,
$h \geq x+1$.
We must use the sign $+$.
\item
Case (3):
$-4<x<-\frac{3}{2}$, $\delta<0$, $h>x+5+\frac{6}{x}$.
We must use the sign $+$.
\item
Case (4):
$x>0$, $\delta<0$, $x+5+\frac{6}{x}>h \geq 1+x$.
We must use the sign $-$.
\item
Case (5):
$-\frac{3}{2} \leq x<0$, $\delta>0$, $h \geq x+1$.
We must use the sign $-$.
\item
Case (6):
$-4<x<-\frac{3}{2}$, $\delta>0$, $h>x+5+\frac{6}{x}$.
We must use the sign $-$.
\item
Case (7):
$x=-4$, $\delta>0$. We must use the sign $-$.
\item
Case (8):
$x=-4$, $\delta<0$. We must use the sign $+$.
\end{itemize}
The solution of $\mathcal{F}(R,G)$ is given by
\begin{equation}
\mathcal{F}(R,G)=\frac{\alpha}{(f(R,G))^{x+2}}R+\frac{\delta}{(f(R,G))^{x}}G
-\frac{6h}{(f(R,G))^{x+4}}\left[4h^{2}\delta x+\alpha(x+2+h)\right]\,,
\label{funzione}
\end{equation}
where
$f(R,G)$ is given by Eq.~(\ref{radice}).
This is an exact solution of EOM in Eqs.~(\ref{un}) and (\ref{dos})
for the Big Rip case.

We show several examples.
In the case $\alpha=1$ and $x=-2$, we find
\begin{equation}
\mathcal{F}(R,G)=R+\dfrac{\sqrt{6}\sqrt{h(1+h)}}{(1-h)}\sqrt{G}\,,
\quad
h\neq 1\,,
\end{equation}
which is in agreement with the result of the previous section.

If $\alpha=0$ and $x=h-3$
(this case corresponds to the cases (1)--(6) presented above),
we find
\begin{equation}
\mathcal{F}(R,G)=\delta G^{\frac{h+1}{4}}\,,
\quad
\delta \neq 0\,,
\end{equation}
which is equivalent to
Eq.~(\ref{trivial}) with $\alpha_{1}=\alpha_{2}=0$.

If $x=-4$, the result is given by
\begin{equation}
\mathcal{F}(R,G)=
\dfrac{16h^{4}\delta}{(1+2h^{2})^{2}}\left[(9+21h+6h^{2})-(1+h)^{2}
\frac{R^{2}}{G}\right]\,,
\quad
\delta \neq 0\,.
\label{Dante}
\end{equation}
Hence,
if for large values of R and G,
$\mathcal{F}(R,G)\sim\pm\alpha\mp\delta(R^{2}/G)$ with
$\alpha>0$ and $\delta>0$, the Big Rip singularity could appear.

If $x=h-1$, the solution
becomes (by absorbing some constant)
\begin{equation}
\mathcal{F}(R,G)=
\delta G \left( \dfrac{R}{G} \right)^{\frac{1-h}{2}}\,,
\quad
\delta \neq 0\,,
\quad
h \neq 1\,.
\label{Sturmtruppen}
\end{equation}
Thus,
if for large values of $R$ and $G$,
$\mathcal{F}(R,G)\sim \delta G^{\gamma}/R^{\gamma-1}$
with $\delta \neq 0$ and $1/2<\gamma<1$ or $1<\gamma<+\infty$,
the Big Rip singularity could appear.

Furthermore,
it is possible to verify that the model:
\begin{equation}
\mathcal{F}(R,G)=\gamma\frac{G^{m}}{R^{n}}\,,
\label{Kriegsmarine}
\end{equation}
with $\gamma$ being a generic constant,
is a solution of Eqs.~(\ref{un}) and (\ref{dos})
in the case of
the Big Rip singularity ($\beta=1$) for some value of $h$. In general, it is
possible to obtain solutions for $h>0$ if $m>0$, $n>0$ and $m>n$. For
example,
the case $n=2$ and $m=3$
realizes the singularity in $h=5$;
the case $n=1$ and $m=3$ realizes the singularity in $h=4+\sqrt{19}$
and so forth.
This is a generalization of Eq.~(\ref{Sturmtruppen}).
Note that we do not recover a
physical solution for $m=-1$ and $n=-2$ because
in this case $h=-3$. For a similar kind of model $F(R^{2}/G)$
which produces the Big Rip singularity, see Eq.~(\ref{Dante}).
For $m=0$ or $n=0$, we recover Eq.~(\ref{trivial}).

\subsubsection{Other types of singularities}

Next, we study the other types of singularities.
We consider the case in which $H$ is given by
\begin{equation}
H=\frac{h}{(t_{0}-t)^{\beta}}\,.
\end{equation}
An exact solution of Eq.~(\ref{Pi}) is given by
\begin{eqnarray}
P(t) \Eqn{=} -\lambda(4h^{2})(t_{0}-t)\,, \\
Z(t) \Eqn{=} \lambda(t_{0}-t)^{2\beta+1}\,,
\end{eqnarray}
where $\lambda$ is a generic constant. The form of $Q(t)$ is given by
\begin{equation}
Q(t)=\dfrac{24h^{4}\lambda}{(t_{0}-t)^{2\beta-1}}+\dfrac{48h^{3}\beta}{(t_{0
}-t)^{\beta}}\,.
\end{equation}
For $\beta=1$, we find a special case of Eq.~(\ref{funzione}). For
$\beta>1$,
we obtain the asymptotic real solution of Eq.~(\ref{t}):
\begin{equation}
t_{0}-t
=f(R,G)=2^{1/2\beta}\left[\dfrac{h^{2}R+\sqrt{h^{4}R^{2}+6h^{4}(4\beta^{2}-1
)G}}{(1+2\beta)G}\right]^{1/2\beta}\,.
\end{equation}
The form of $\mathcal{F}(R,G)$ is expressed as
\begin{equation}
\mathcal{F}(R,G)=-4h^{2}\lambda (f(R,G))R+\lambda
(f(R,G)^{1+2\beta})G+24h^{4}\lambda (f(R,G)^{1-2\beta})\,,
\quad
\beta>1\,.
\end{equation}
This is an asymptotic solution of Eq.~(\ref{prime})
when (for $\beta>1$)
\begin{equation}
-\frac{1}{\kappa^{2}}\left[2\dot H +3(1+w)H^{2}\right]\sim
-\dfrac{3(1+w)h^{2}}{\kappa^{2}}(t_{0}-t)^{-2\beta}\,.
\label{auto}
\end{equation}
In the case $\beta \gg 1$, the form of $\mathcal{F}(R,G)$ is written as
\begin{equation}
\mathcal{F}(R,G)
\simeq
\lambda\left(\dfrac{\alpha G}{R+\sqrt{R^{2}+\gamma G}}-R
\right)\,,
\quad
\alpha>0\,,
\quad
\gamma>0\,,
\quad
\lambda \neq 0\,.
\end{equation}
This is the asymptotic behavior of a $\mathcal{F}(R,G)$ model in which
a ``strong'' Type I singularity ($\beta \gg 1$) could appear.
By taking $g(R,G)=\gamma G^{m}/R^{n+1}$ and using
Eqs.~(\ref{tic}) and (\ref{tac}), it is possible to verify that for the
model
\begin{equation}
\mathcal{F}(R,G)=\gamma\frac{G^{m}}{R^{n}}\,,
\label{miseriaccia}
\end{equation}
the function $\mathcal{G}(H, \dot{H}...)$ in Eq.~(\ref{Feldwebel}) is
given by
\begin{equation}
\mathcal{G}(H,\dot{H}...)\simeq
-\frac{3 h^{2}(2m-n-1)(1+w)}{\kappa^{2}(t_{0}-t)^{2\beta}}\,,
\end{equation}
which is, under the condition $2m-n-1>0$, an
asymptotic solution of Eq.~(\ref{prime}) in the case of $\beta>1$.
Thus,
in the model $\mathcal{F}(R,G)\simeq \gamma G^{m}/R^{n}$ with $m>(n+1)/2$
the Type I singularity could appear.
This point has important consequences because it is possible to see that the
theories $F(R)=R^{n}$ with $n>1$ or $F(G)=G^{m}$ with $m>1/2$ can become
singular.

To find other models, we can consider the results of Sec.~III.
The Type I singularities correspond to the asymptotic limits for $R$ and $G$
\begin{equation}
R\sim 12 H^{2}\,,
\quad
G\sim 24 H^{4}\,.
\end{equation}
These are two functions of the Hubble parameter only, so that
\begin{equation}
\lim_{t\rightarrow t_{0}} 24\left( \frac{R}{12} \right)^{2}
= \lim_{t\rightarrow t_{0}} G\,.
\label{Limite}
\end{equation}
If we substitute $G$ for $R$ in Eq.~(\ref{primo}) by taking into account
Eq.~(\ref{Limite}), we obtain a
zero function (this is because Eq.~(\ref{primo}) is zero on the singularity
solution).
If we substitute $G$ for $G/R$, however, we obtain the following model:
\begin{equation}
\mathcal{F}(R,G)= R-\dfrac{6G}{R}\,.
\label{zap}
\end{equation}
This is an asymptotic solution of Eq.~(\ref{prime}) such as
Eq.~(\ref{auto}).
Thus, there appears Type I singularity for
$\mathcal{F}(R,G)\sim R-\alpha(G/R)$ with $\alpha>0$.

In the case of $H=h/(t_{0}-t)^{\beta}$ with $\beta<1$, it is not possible to
write $G$ and $R$ like functions of the same variable ($H$ or the same
combination of $H$ and $\dot H$). Nevertheless, if we
examine the asymptotic behavior of $G$ and $R$, we have
\begin{eqnarray}
R \Eqn{\simeq}
\frac{6h\beta}{(t_{0}-t)^{\beta+1}}\,, \\
G \Eqn{\simeq}
\frac{24 h^{3}\beta}{(t_{0}-t)^{3\beta+1}}\,,
\end{eqnarray}
and
\begin{equation}
\frac{G}{R}\sim G^{\frac{2\beta}{3\beta+1}}\label{Todt}
\end{equation}
If we use $G/R$ for $G$ in Eq.~(\ref{secondo}) as in Eq.~(\ref{Todt}), we
see that the asymptotic time dependence in Eq.~(\ref{prime}) for $\beta<1$
is
the same:
\begin{equation}
-\frac{1}{\kappa^{2}}\left[2\dot H +3(1+w)H^{2}\right]\sim
\frac{\alpha}{(t_{0}-t)^{\beta+1}}+\frac{\gamma}{(t_{0}-t)^{2\beta}}\,.
\end{equation}
Under this consideration, it is possible to derive
a $\mathcal{F}(R,G)$-gravity theory (by setting some parameters) from
Eq.~(\ref{secondo}) as
\begin{equation}
\mathcal{F}(R,G)=R+\frac{3}{2}\frac{G}{R}\,,
\end{equation}
in which the other types of singularities appear.
Thus, in this model ($\mathcal{F}(R,G)\sim R+\alpha(G/R)$ with $\alpha>0$)
the Type II, III and IV singularities could appear.
Then, by substituting $G$ for $R$ we get
\begin{equation}
\mathcal{F}(R,G)
\simeq
R-\delta\frac{(1+\beta)}{(\beta-1)}
|R|^{\frac{2\beta}{1+\beta}}\,,
\quad
\delta>0\,.
\end{equation}
This is a well-know result. In the model
$\mathcal{F}(R\rightarrow\infty)\sim R+\alpha R^{\gamma}$,
for $0<\gamma<1$ and $\alpha>0$,
a Type III singularity could appear.
In the model $\mathcal{F}(R\rightarrow -\infty)\sim R+\alpha |R|^{\gamma}$,
for $-\infty<\gamma<0$ and $\alpha>0$,
a Type II singularity could appear.
In the model $\mathcal{F}(R\rightarrow 0^{-})\sim R+\alpha |R|^{\gamma}$,
for $2<\gamma<+\infty$ ($\gamma\neq 2n/(n-1)$, where $n$ is a natural
number)
and $\alpha<0$, a Type IV singularity could appear.
(In the Big Rip case, we
have found exact solutions. This kind of reasoning is therefore
inapplicable.)

\subsection{Realistic model of $\mathcal{F}(R,G)$-gravity}

We study the following realistic model of $\mathcal{F}(R,G)$-gravity:
\begin{equation}
\mathcal{F}(R,G)= a_{1}G^{n}+a_{2}R^{m}
+ \frac{a_{5}}{a_{3}G^{n}+a_{4}R^{m}}\,,
\label{GutenAppetit}
\end{equation}
where $a_{i} (i=1, \cdots, 5)$ are constants and $n ( > 0)$ and $m ( > 0)$
are positive constants.
For large values of $R$ and $G$, we have
\begin{equation}
\mathcal{F}(R,G)\simeq a_{1}G^{n}+a_{2}R^{m}\,.
\end{equation}
In the specific case $n \geq 3$ and $m=(1/2)(1+2n+\sqrt{3-12n+4n^{2}})$
(in which, $m \geq \left(7+\sqrt{3}\right)/2$),
from Eq.~(\ref{trivial}) we see that the Big Rip singularity could occur.
To find other singularity solutions, we investigate the asymptotic form of
$\mathcal{G}(H,\dot{H}...)$ in Eq.~(\ref{Feldwebel})
and require the consistence with Eq.~(\ref{prime})
of Eq.~(\ref{zuzzurellone}). The behavior
of Eq.~(\ref{Feldwebel}) takes two different asymptotic forms
which depend on the parameter of $\beta$ as follows:
\begin{itemize}
\item
Case of $\beta \geq 1$:
In the limit $t\rightarrow t_{0}$, we find
\begin{eqnarray}
\mathcal{G}(H, \dot{H}...) \Eqn{\sim} \frac{1}{g(R,G)} \biggl\{ \alpha
\left[ \mathcal{F}+\frac{{\mathcal{F}}'_{R}}{(t_{0}-t)^{2\beta}}+
\frac{{\mathcal{F}}'_{G}}{(t_{0}-t)^{4\beta}} \right]
\nonumber \\
&&
+\gamma \left[ \frac{1}{(t_{0}-t)^{2\beta}}\frac{dg(R,G)}{dR}
+\frac{df(R,G)}{dR} \right] \frac{1}{(t_{0}-t)^{2\beta}}
\nonumber \\
& &
+\delta\frac{{\dot{\mathcal{F}}}'_{R}}{(t_{0}-t)^{\beta}}
+\epsilon\frac{{\dot{\mathcal{F}}}'_{G}}{(t_{0}-t)^{3\beta}}
+\zeta{\ddot{\mathcal{F}}}'_{R}
+\eta\frac{{\ddot{\mathcal{F}}}'_{G}}{(t_{0}-t)^{2\beta}} \biggr\}\,,
\label{Achtung}
\end{eqnarray}
where $\epsilon$ and $\eta$ are constants.
To realize a I Type singularity, from Eq.~(\ref{prime}) we must have
\begin{equation}
\mathcal{G}(H, \dot{H}...)\sim
-\frac{3(1+w)h^{2}}{\kappa^{2}(t_{0}-t)^{2\beta}}\,.
\label{AirOne}
\end{equation}
Hence,
if for $G\sim 1/(t_{0}-t)^{4\beta}$ and $R\sim 1/(t_{0}-t)^{2\beta}$
with $\beta \geq 1$, the highest term of Eq.~($\ref{Achtung}$) is
proportional to $1/(t_{0}-t)^{2\beta}$, it is possible to have a Type I
singularity.
As in $F(G)$-gravity, this condition is
necessary and not sufficient.
\item
Case of $\beta<1$:
In the limit $t\rightarrow t_{0}$, we obtain
\begin{eqnarray}
\mathcal{G}(H, \dot{H}...) \Eqn{\sim}
\frac{1}{g(R,G)} \biggl\{ \alpha \left[ \mathcal{F}
+\frac{{\mathcal{F}}'_{R}}{(t_{0}-t)^{\beta+1}}
+\frac{{\mathcal{F}}'_{G}}{(t_{0}-t)^{3\beta+1}} \right] + \gamma \biggl[
\frac{1}{(t_{0}-t)^{\beta+1}}\frac{dg(R,G)}{dR}
\nonumber \\
& & \hspace{-20mm}
+\frac{df(R,G)}{dR} \biggr] \frac{1}{(t_{0}-t)^{\beta+1}}
+\delta\frac{{\dot{\mathcal{F}}}'_{R}}{(t_{0}-t)^{\beta}}
+\epsilon\frac{{\dot{\mathcal{F}}}'_{G}}{(t_{0}-t)^{2\beta+1}}
+\zeta{\ddot{\mathcal{F}}}'_{R}
+\eta\frac{{\ddot{\mathcal{F}}}'_{G}}{(t_{0}-t)^{2\beta}} \biggr\}\,.
\label{Hilfe}
\end{eqnarray}
To realize this kind of singularities, from Eq.~(\ref{prime})
we must have
\begin{equation}
\mathcal{G}(H, \dot{H}...)\sim
-\frac{2\beta h}{\kappa^{2}(t_{0}-t)^{\beta+1}}\,.
\label{Panzer}
\end{equation}
Thus,
if for $G\sim 1/(t_{0}-t)^{3\beta+1}$ and $R\sim 1/(t_{0}-t)^{\beta+1}$
with $\beta<1$, the highest term of Eq.~($\ref{Hilfe}$) is proportional to
$1/(t_{0}-t)^{\beta+1}$, it is possible to have a Type II, III or IV
singularity. Also this condition is necessary and not sufficient.
\end{itemize}

For the model in Eq.~(\ref{GutenAppetit}),
if $m$ and $n$ are positive numbers
($\mathcal{F}(R\rightarrow\infty, G\rightarrow\infty)
\simeq a_{1}G^{n}+a_{2}R^{m}$) and
$m=2n$, the asymptotic behavior of Eq.~(\ref{Achtung}) (in this case
$g(R,G)=a_{2}R^{m-1}$) when
$R \simeq 12h^{2}/(t_{0}-t)^{2\beta}$
and
$G \simeq 24 h^{4}/(t_{0}-t)^{4\beta}$ is proportional to
$1/(t_{0}-t)^{2\beta}$ and therefore it is possible to realize the Type I
singularity.
As a consequence, we get
\begin{equation}
\mathcal{G}(H,\dot{H}...)
\simeq
-\frac{3(1+w)h^{2}}{\kappa(t_{0}-t)^{2\beta}}
\left[ m-1+\frac{(m-2)a_{1}}{a_{2}} \right]\,.
\end{equation}
To have the consistence with Eq.~(\ref{AirOne}), we find that if
$1 \leq m<2$ and $a_{1}/a_{2}<0$,
the Type I singularity could appear
(for example, if $m=1$, $n=1/2$, $a_{2}=1$ and $a_{1}<0$, we recover the
case
of Eq.~(\ref{primo})). The same result is obtained
if $m>2$ and $a_{1}/a_{2}>0$.

We see that for $m>0$, $n>0$ and $-1<\beta<1$,
${\ddot{\mathcal{F}}}'_{R}/g(R,G)\sim (t_{0}-t)^{-2}$
(like above, $g(R,G)=a_{2}R^{m-1}$)
and $\mathcal{G}(H, \dot{H}...)$ in Eq.~(\ref{Hilfe}) diverges faster than
$(t_{0}-t)^{-\beta-1}$, so that in order to find Type II or III
singularities,
we must have ${\ddot{\mathcal{F}}}'_{R}=0$. In general, this is true if
$m=1$
and we can recover the results in Sec.~III for $F(G)$-gravity.

When $R\rightarrow0^{-}$ and $G\rightarrow0^{-}$, for $m>0$ and $n>0$,
$\mathcal{F}(R,G)$ behaves as
\begin{equation}
\mathcal{F}(R\rightarrow0^{-}, G\rightarrow0^{-})
\simeq\frac{a_{5}}{a_{3}G^{n}+a_{4}R^{m}}\,.
\end{equation}
In this case, if $\beta<-1$, Eq.~(\ref{Hilfe})
($g(R,G)=a_{5}/(a_{3}G^{n}R+a_{4}R^{m+1})$) diverges and Eq.~(\ref{Panzer})
becomes inconsistent, so that the model is free of Type IV singularities.

\section{Curing the finite-time future singularities}

In this section, we discuss a possible way to cure the finite-time future
singularities in $F(G)$-gravity and $\mathcal{F}(R,G)$-gravity.
In the limit of large curvature, the quantum effects become important
and lead to higher-order curvature corrections.
It is therefore interesting to resolve the finite-time future singularities
with some power function of $G$ or $R$.

\subsection{$F(G)$-gravity}

First, we consider $F(G)$-gravity.
If any singularity occurs, Eq.~(\ref{prime}) behaves as
\begin{equation}
\mathcal{G}(H, \dot{H}...)\simeq
\begin{cases}
-\frac{3(1+w)h^{2}+2\beta h}{\kappa^{2}}(t_{0}-t)^{-2} &
\mbox{Big Rip}\\
-\frac{3(1+w)h^{2}}{\kappa^{2}}(t_{0}-t)^{-2\beta} &
\mbox{$\beta>1$ (Type I)}\\
-\frac{2\beta h}{\kappa^{2}}(t_{0}-t)^{-\beta-1} &
\mbox{$\beta<1$ (Types II, III, IV )}
\end{cases}
\label{casistiche}
\end{equation}
The singularities appear in two cases: $G\rightarrow \pm\infty$ or
$G\rightarrow 0^{-}$.

\vspace{2mm}
\noindent
{\bf (i) {\em Case of} $G\rightarrow \pm\infty$}
\vspace{1mm}

Suppose that for large values of $G$,
\begin{equation}
R+F(G\rightarrow \pm\infty)\longrightarrow R+\gamma G^{m}\,,
\quad
m\neq 1
\,,
\label{eins}
\end{equation}
with $\gamma \neq 0$.
One way to prevent a singularity appearing could be
that the function $\mathcal{G}(H, \dot H...)$ becomes inconsistent with the
behavior of
Eq.~(\ref{casistiche}). In general, $\mathcal{G}(H, \dot H...)$ must tend to
infinity faster than Eq.~(\ref{casistiche}). For
$H=h/(t_{0}-t)$
(Big Rip), we have
\begin{equation}
\mathcal{G}(H, \dot H...)\simeq \frac{\alpha}{(t_{0}-t)^{4m}}\,.
\end{equation}
Hence, if $m > 1/2$, we avoid the singularity. Nevertheless, there is one
specific case in which the Big Rip singularity could occur. If $m=(1+h)/4$,
$\mathcal{G}(H, \dot H...)$ is exactly equal to zero, so that (for example)
the following specific model admits the Big Rip singularity:
\begin{equation}
R+F(G)=R+\frac{\sqrt{24 m (4m-1)^{3}}}{2h(1-2m)}G^{1/2}+\gamma G^{m}\,.
\label{esempio}
\end{equation}
This is because the power function $G^{m}$ is an invariant
with respect to the Big Rip singularity of $G^{1/2}$.
If for large
values of $G$, $F(G)\sim \alpha G^{1/2}$, we can eliminate the Big Rip
singularity with a power function $\gamma G^{m}$ ($m \geq 2$) only if
$\alpha>0$.

For $H=h/(t_{0}-t)^{\beta}$ with $\beta>1$ (Type I) and
the behavior in Eq.~(\ref{eins}), we find
\begin{equation}
\mathcal{G}(H, \dot H...)\simeq \dfrac{\alpha}{(t_{0}-t)^{4\beta m}}\,.
\end{equation}
Also in this case, if $m>1/2$, we avoid the singularity.
For example,
$R+F(G)=R+\alpha\sqrt{G}+\gamma G^{2}$ with $\alpha>0$ is free of
Type I singularities,
while if $\alpha<0$,
the Big Rip singularity could appear.

For $H=h/(t_{0}-t)^{\beta}$ with $0<\beta<1$
(Type III) and the behavior in Eq.~(\ref{eins}), we obtain
\begin{equation}
\mathcal{G}(H, \dot H...)\simeq
\frac{\alpha}{(t_{0}-t)^{m(3\beta+1)+(1-\beta)}}\,.
\end{equation}
If $m> 2\beta/(3\beta+1)$ (i.e. $m> 1/2$), we avoid the singularity.

Also for $H=h/(t_{0}-t)^{\beta}$ with $-1/3<\beta<0$
(Type II, $G\rightarrow-\infty$), we have to require the same condition.
For example,
$R+\alpha |G|^{\zeta}+\gamma G^{2}$ with $\zeta<1/2$ is free of
Type I, II and III singularities.

\vspace{2mm}
\noindent
{\bf (ii) {\em Case of} $G\rightarrow 0^{-}$}
\vspace{1mm}

Suppose that for small values of $G$,
\begin{equation}
R+F(G\rightarrow 0^{-})\longrightarrow R+\gamma G^{m}\,,
\label{zwei}
\end{equation}
with $\gamma \neq 0$ and $m$ being an integer.
For $H=h/(t_{0}-t)^{\beta}$ with $\beta<-1/3$
(Type II and IV singularities), we get
\begin{equation}
\mathcal{G}(H, \dot H...)\simeq
\frac{\alpha}{(t_{0}-t)^{m(3\beta+1)+(1-\beta)}}\,,
\end{equation}
which diverges and hence becomes inconsistent with Eq.~(\ref{casistiche})
if $m< 2/3$.
For example,
$\alpha |G|^{\zeta}+\gamma G^{-1}$ with $\zeta>2/3$ is free of Types IV
singularities.

As a result,
the term $\gamma G^{m}$ with $m> 1/2$ and $m\neq 1$ cure the singularities
occurring when $G\rightarrow \pm\infty$.
Moreover, the term $\gamma G^{m}$ with $m \leq 0$ and $m$ being
an integer cure the singularities occurring when $G\rightarrow 0^{-}$.

In $f(R)$-gravity, 
by using the term $\gamma R^{m}$, the same consequences are found. 
The term $\gamma R^{m}$ with $m>1$ cures the Type II and III 
singularities. 
On the other hand, the term $\gamma R^{m}$ with $m<2$ cures the Type IV
singularity.

Note that $\gamma G^{m}$ or $\gamma R^{m}$ are invariants
with respect to the Big Rip solution (see Eq.~(\ref{trivial})), so it is
necessary to pay attention to the whole form of the
$F(G)$ or $f(R)$-gravity (see Eq.~(\ref{esempio})).

It is also
possible to cure the singularities in a $F(G)$-gravity theory
with the power functions of $R$ and
a $f(R)$-gravity theory with the power functions of $G$.
To do it,
it is useful to take into account that $G$ diverges as $R^{2}$ in the Type I
singularity solutions, at least as $R$ in the Type II, and as $R^{2}$
in the Type III, and $G$ tends to zero at least as $R^{3}$ in the Type IV
(proved by the fact that when $\beta<1$, $G\sim R^{3\beta+1/\beta+1}$).
We show several examples.
\begin{itemize}
\item
$R+G^{n}+R^{m}\sim R+R^{m}$ on the asymptotic limit of the Types I and III
singularity solutions (if they exist) when $m>2n$.
\item
$R+G^{n}+R^{m}\sim R+R^{m}$ on the Types II singularity solutions if $m>n$
(on
this kind of solutions $R$ tends always to infinity,
while in some cases, for $-1<\beta<-1/3$, G tends to zero).
\item
$1/(G^{n}+R^{m})\sim 1/R^{m}$ on the Types IV singularity solutions (for
which
$R\rightarrow0^{-}$ and $G\rightarrow0^{-}$) if $m<3n$.
\end{itemize}
Thus, the singularity solutions found in
Sec.~III for $F(G)$-gravity can be cured by the term $\gamma R^{m}$ with
$m>1$
for Type I, II and III singularities and
that with $m<1$ for Type IV singularity.

We mention the Type I singularities with $\beta>1$.
We have shown that the model $R+\gamma G^{m}$ with $m>1/2$ ($m\neq 1$) 
is free of Type I singularity because 
Eq.~(\ref{prime}) becomes inconsistent. Nevertheless,
it follows from Eq.~(\ref{miseriaccia}) that
the models $G^{m}$ with
$m>1/2$ and $R^{n}$ with $n>1$ can show the Type I singularity
in the asymptotic limit.
This means that when $t$ is very close to $t_{0}$,
the term $G^{m}$ or $R^{n}$ is dominant
over $R$, $R+G^{m}\sim G^{m}$
(or $R+R^{n}\sim R^{n}$) and therefore
the Type I singularity could appear.
Hence,
the important point is whether the model can approach to very large values
of
$R$ or $G$ with a non-singular metric (which is not admitted) and
then become singular because $R$ is negligible.
It depends on the form and the
dynamics of the model and the value of $m$ or $n$.
If $m \gg 1/2$ or $n \gg 1$,
the singularity could appear more easily.
Thus, in order to avoid the singularity solutions,
it is better to choose $m$ and $n$ as $m>1/2$ and
as $n\gtrsim 1$, respectively, but $m$ and $n$ are not very large.
This does not hold
in the other types of singularities. The theory
$R+G^{m}$ with $m\leq 0$ and $m$ being an integer (or $R+R^{n}$ with $n<2$)
is free of Type II, III and IV singularities
as the theories $G^{m}$ or $R^{n}$.

\subsection{$\mathcal{F}(R,G)$-gravity}

Next, we study $\mathcal{F}(R,G)$-gravity.
In the general $\mathcal{F}(R,G)$-gravity, in order to avoid the
singularities with power functions, we must require that the EOM
(\ref{un}) and (\ref{dos}) are
inconsistent on the singularities solutions.
Within the framework of $\mathcal{F}(R,G)$-gravity, we can use the terms
such as $G^{m}/R^{n}$ to cure the singularities.
The singularities appear in the following three cases:
(a)
$R\rightarrow \pm\infty$, $G\rightarrow \pm\infty$ (Types I, II, III),
(b)
$R\rightarrow -\infty$,
$G\rightarrow 0^{-}$
(Type II for $-1<\beta<-1/3$),
and
(c)
$R\rightarrow 0^{-}$, $G\rightarrow 0^{-}$ (Type IV).

We investigate general cases.
Suppose that for large values of $G$ and $R$,
\begin{equation}
\mathcal{F}(R\rightarrow \infty, G\rightarrow \infty)\longrightarrow R
+\gamma\frac{G^{m}}{R^{n}}\,,
\label{R+F}
\end{equation}
with $\gamma \neq 0$.
In the case of the Big Rip singularity,
in which $H$ is given by Eq.~(\ref{H}),
$\mathcal{G}(H,\dot{H}...)$ in Eq.~(\ref{Feldwebel})
diverges as
\begin{equation}
\mathcal{G}(H,\dot{H}...)\sim
\frac{\alpha}{(t_{0}-t)^{4m-2n}}\,.
\end{equation}
Thus, if $m > (n+1)/2$, we avoid the singularity. Nevertheless,
there is the possibility that
$\mathcal{G}(H, \dot{H}...)$ is exactly equal to zero and the
Big Rip singularity could occur
(see Eqs.~(\ref{Sturmtruppen}) and (\ref{Kriegsmarine})
in the case of $m=n+1$).
Hence, the whole form of $\mathcal{F}(R,G)$
as well as its form in the asymptotic limit must be examined.

In the case of Eq.~(\ref{HII})
(Type I), $\mathcal{G}(H,\dot{H}...)$ diverges as
\begin{equation}
\mathcal{G}(H,\dot{H}...)\sim
\frac{\alpha}{(t_{0}-t)^{4\beta m-2\beta n}}\,.
\end{equation}
Also in this case, if $m>(n+1)/2$, we avoid the singularity.
Similarly to the above, however,
if $m \gg 1$ and $n \ll 1$,
the asymptotic limit of $\mathcal{F}(R,G)$ in Eq.~(\ref{R+F}) behaves as
$\gamma G^{m}/R^{n}$ and therefore
the Type I singularity could occur (see Eq.~(\ref{miseriaccia})).

As a consequence,
we can avoid the Type I singularities if the asymptotic
behavior of the model is given by Eq.~(\ref{R+F}) and
its asymptotic form has the power functions
\begin{equation}
G^{m}R^{n}\,,
\end{equation}
or
\begin{equation}
\frac{G^{m}}{R^{n}}\,,
\quad
m>\frac{n+1}{2}\,,
\end{equation}
with $m$ and $n$ being positive integers.

Now, suppose that
when $H=h/(t_{0}-t)^{\beta}$
with $\beta<1$, the asymptotic limit of $\mathcal{F}(R,G)$ becomes
\begin{equation}
\mathcal{F}(R,G)\longrightarrow \gamma\frac{G^{m}}{R^{n}}\,.
\end{equation}
For $\beta<1$, $\mathcal{G}(H, \dot{H}...)$ behaves as
\begin{equation}
\mathcal{G}(H,\dot{H}...)\sim\frac{\alpha}{(t_{0}-t)^{2}}\,,
\label{same}
\end{equation}
which diverges faster than $(t_{0}-t)^{-\beta-1}$ and therefore
the Type II, III and IV singularities are always avoided for any value of
$m$ and $n$.
The same scenario to cure the future singularity by adding the non-singular
modified gravity maybe applied here again.

\section{Conclusion}

In the present paper, we have investigated the
finite-time future singularities in $F(G)$-gravity and
$\mathcal{F}(R,G)$-gravity.
We have reconstructed the $F(G)$-gravity and
$\mathcal{F}(R,G)$-gravity models in which
the finite-time future singularities may occur.
It has been demonstrated that all four types of finite-time future
singularity
may emerge for a variety of the above models with the effective
quintessence/phantom EoS behavior in the same qualitative way as for
convenient DEs where also all four types of future singularity may
occur~\cite{Nojiri:2005sx, sing1, sing2}.
This provides the explicit demonstration that whatever the effective DE
model (convenient one or modified gravity) is, it may lead to singular
future
universe. Moreover, the future singularity may manifest itself as radius
singularity for spherically-symmetric spaces. This may cause instabilities
for black holes~\cite{sing1} and relativistic
stars~\cite{Kobayashi:2008tq, Kobayashi:2008wc,
Babichev:2009td, Upadhye:2009kt}. Other imprints of the
singular future universe to current cosmology may be searched as well.

However, there exists fundamental qualitative difference between convenient
DE and modified gravity. It turns out that sometimes it is possible to solve
the singularity issue taking account of quantum gravity effects (see Big
Rip singularity resolution in Ref.~\cite{elizalde}) or by the coupling of DE
with Dark Matter (DM)
(some fine-tuning of initial conditions may help to resolve Type II or Type
IV future singularity~\cite{DM}). Nevertheless, quantum gravity account is
effectively the modification of gravity. Moreover, it is only modified
gravity (actually, its additional modification as we have demonstrated on
the
example of non-singular $F(G)$-model in Subsection III. C) may suggest the
universal scenario to cure any finite-time future singularity. This is
achieved by adding such non-singular theory to any DE containing future
singularity in its evolution. Furthermore, such additional modification
may always be made by terms which are relevant only in the early universe
and
are typical as quantum gravity corrections.
Hence, modified gravity suggests the universal scenario to protect the
future universe from singularity while not destroying the attractive
cosmological properties of specific DE alternative gravity like its
viability with local/cosmological tests if exists. This may be considered as
powerful theoretical argument in favor of the consideration of such theories
as DEs.

\section*{Acknowledgments}
K.B. and S.D.O. thank Professor Shin'ichi Nojiri for his
collaboration in the previous work~\cite{Odintsov}.
The work is supported in part by
the National Science Council of R.O.C. under
Grant \#s: NSC-95-2112-M-007-059-MY3 and NSC-98-2112-M-007-008-MY3 and
National Tsing Hua University under the Boost Program and Grant \#:
97N2309F1 (K.B.);
MEC (Spain) project FIS2006-02842 and AGAUR (Catalonia) 2009SGR-994 and by
JSPS Visitor Program (Japan) (S.D.O.) and by INFN (Trento)-CSIC (Barcelona)
exchange grant.

\appendix

\section{Simple model of $f(R)$-gravity}

There is a great diffusion of modified gravity models which for large
values of curvature tend to a constant and imitate the $\Lambda$CDM model.
Nevertheless, the existence of derivatives of the modified
function, which is very small but different
from zero, involves the possibility of singularities for $\beta<1$.
In this Appendix, we explore the finite-time future singularities
in a simple model of $f(R)$-gravity.
The action of $f(R)$-gravity is given by
\begin{equation}
S=\int d^{4}x \sqrt{-g} \left[ \frac{1}{2\kappa^2}
\left( R+f(R) \right)
+{\mathcal{L}}_{\mathrm{matter}}
\right]\,,
\label{eq:f(R)}
\end{equation}
which corresponds to the action in Eq.~(\ref{azione}) with
$\mathcal{F}(R,G) = R+f(R)$.
We examine the following simple $f(R)$-gravity model~\cite{Cognola:2007zu}
which reproduces the current accelerated expansion of the universe
and imitate a cosmological constant for large values of curvature:
\begin{equation}
f(R)=\alpha (e^{-b R}-1)\,,
\label{generic}
\end{equation}
where $b\, ( > 0 )$ is a positive constant.
In what follows, we consider the pure gravitational action of $f(R)$-gravity
in Eq.~(\ref{eq:f(R)}) without ${\mathcal{L}}_{\mathrm{matter}}$.

The finite-time future singularities in $f(R)$-gravity
have been discussed in Ref.~\cite{Odintsov},
from which we propose the asymptotic expression of $\mathcal{G}(H,\dot
H,...)$
when $\beta<1$, similarly to Eq.~(\ref{ja}):
\begin{equation}
\mathcal{G}(H,\dot H,...)\sim
\alpha
f(R)+\frac{\gamma}{(t_{0}-t)^{\beta+1}}f'(R)+\frac{\delta}{(t_{0}-t)^{\beta+
3}}f''(R)+\frac{\zeta}{(t_{0}-t)^{2\beta+4}}f'''(R)\,.
\label{VonRichthoffen}
\end{equation}
Here, the prime denotes differentiation with respect to $R$.
When $\beta<0$, the asymptotic behavior of $R$ is given by
\begin{equation}
R\sim 6h\beta(t_{0}-t)^{-\beta-1}\,.
\end{equation}
We assume $R>0$, so in this case $h$ must be negative.
If $H$ behaves as
\begin{equation}
H\sim h(t_{0}-t)^{-\beta}+H_{0}\,,
\end{equation}
as Eq.~(\ref{Hsingular}),
$H$ can be still positive in the limit $t\rightarrow t_{0}$.
For large values of $R$ ($-1<\beta<1$),
$f(R)$ in Eq.~(\ref{generic}) tend to $-\alpha$ and
\begin{equation}
f'(R)\sim -\alpha b e^{-b R}\,,
\quad
f''(R)\sim \alpha b^{2} e^{-b R}\,,
\quad
f'''(R)\sim -\alpha b^{3} e^{-b R}\,.
\end{equation}
To obtain a singularity, the highest term in Eq.~(\ref{VonRichthoffen})
must be divergent as $1/(t_{0}-t)^{\beta+1}$. In order to check it, it is
convenient to develop
the exponential function in power-series. The third
term of Eq.~(\ref{VonRichthoffen}) behaves as
\begin{equation}
\frac{\delta}{(t_{0}-t)^{\beta+3}}f''(R)\sim
\frac{\delta}{\Sigma_{(n=0\rightarrow\infty)}
(t_{0}-t)^{-n(\beta+1)+\beta+3}}\,.
\label{zip}
\end{equation}
For $n=2/(\beta+1)$,
the highest term of the denominator behaves as $(t_{0}-t)^{\beta+1}$.
If $\beta\rightarrow -1^{+}$, $n\rightarrow\infty$ and this
is just the asymptotic value of $n$ of the highest term in Eq.~(\ref{zip}).
A similar argument is valid for the last term of Eq.~(\ref{VonRichthoffen}),
whereas the first and second terms of Eq.~(\ref{VonRichthoffen}) tend to a
constant and zero, respectively.
Thus, a Type II singularity could occur in
the model in Eq.~(\ref{generic}) for large values of $R$.
Note that $f(R)$-gravity unifying the early-time inflation with late-time
acceleration as proposed in Ref.~\cite{2003} 
turns out to be non-singular due to the presence of $R^2$ term.

\section{Asymptotic behavior of singular models}

In this Appendix, we discuss the asymptotic behavior of singular
models. In Sec.~IV A,
we have shown that in principle it is possible to write a general
$\mathcal{F}(R,G)$-gravity theory
in the form in Eq.~(\ref{zuzzurellone}).
In this case,
from Eqs.~(\ref{eq:def-mG}) and (\ref{Feldwebel}) we obtain
\begin{equation}
\mathcal{G}(H,\dot{H}...)=
-\frac{1}{\kappa^{2}} \left[ 2\dot H +3(1+w)H^{2} \right]\,,
\label{b}
\end{equation}
where
\begin{eqnarray}
\mathcal{G}(H,\dot{H}...) \Eqn{=}
\frac{1}{2\kappa^{2}g(R,G)} \biggl\{ (1+w)(\mathcal{F}-R{\mathcal{F}}'_{R}
-G{\mathcal{F}}'_{G})
\nonumber \\
&&
+\left( R\frac{dg(R,G)}{dR}+\frac{df(R,G)}{dR} \right)
\left[ 6H^2(1+w)+4\dot{H} \right]
\nonumber\\
& &
+ H{\dot{\mathcal{F}}}'_{R}(4+6w)
+8H{\dot{\mathcal{F}}}'_{G} \left[ 2\dot{H}+ H^{2}(2+3w) \right]
+2{\ddot{\mathcal{F}}}'_{R}+8 H^{2}{\ddot{\mathcal{F}}}'_{G} \biggr\}\,.
\label{c}
\end{eqnarray}
It is clear that in the case of $\mathcal{F}(R,G)=R+F(G)$,
by taking $g(R,G)=1$ and $f(R,G)=F(G)$ in Eq.~(\ref{zuzzurellone}),
Eqs.~(\ref{prime}) and (\ref{MuyBien}) in Sec.~III A is recovered.

To verify the
existence of singularities, it is very useful to control the consistence of
Eq.~(\ref{b}). On the exact solutions (this is the case of the Big Rip
solutions), this check is independent
of the choice of $g(R,G)$ in Eq.~(\ref{zuzzurellone}).
We must carefully verify only
that $g(R,G)\neq 0$ on the singularity solution (or -equivalently- if we use
this equation to find singularity solutions, it is possible to lost some
solutions in which $g(R,G)= 0$).
When we check asymptotic solutions, a problem could appear.
It is based on
the confrontation between the asymptotic behaviors of the right-hand and
left-hand sides of Eq.~(\ref{b}).
Furthermore, in this case there is
a problem if all the terms of $\mathcal{F}(R,G)$ have the same asymptotic
behavior on the singularity solution. For example, in order to verify
the existence of the Type I singularity solution
in the model $R-\alpha G/R$
with $\alpha>0$ (see Eq.~(\ref{zap})),
it is indifferent to take $g(R,G)=1$ and $f(R,G)=-\alpha G/R$ or
$g(R,G)=-\alpha G/R^{2}$ and $f(R,G)=R$
because on the singularity solution the asymptotic behavior of $R$
($R\sim (t_{0}-t)^{-2}$) is the same
as $G/R$. Nevertheless, if the terms
do not have the same asymptotic behavior (for example,
$\mathcal{F}(R,G)=R+R^{2}+R^{3}...$), it is necessary to be
careful in the choice of $g(R,G)$ to substitute into Eq.~(\ref{c}).
The mechanism is the following:
The right-hand side of Eq.~(\ref{b}) behaves as $R$, while the
left-hand side of Eq.~(\ref{b})
is proportional to $1/g(R,G)$. Automatically, in the asymptotic limit all
the
terms smaller than
$R g(R,G)$ (and also their derivatives) are neglected.
As a consequence,
when we use the EOM in the asymptotic limit, we must choose $g(R,G)$
as the coefficient of the smallest term
which we want to consider.
This is easy to do when the terms are completely
different in the limit. For example,
$R^{2}+1/R \sim R^{2}$ when $R\rightarrow\infty$ (this is the
principle that we have used
in the study of the realistic models shown in the present paper).
The question is trickier when the terms of $\mathcal{F}(R,G)$ tend together
to infinity or to zero with different velocities. In this case, the choice
of
$g(R,G)$ depends on our target, if we want to verify the EOM in
more or less strong limit.
For example, let us consider the model $\mathcal{F}(R,G)=R+R^{2}$. If we
choose $g(R,G)=1$ and $f(R,G)=R^{2}$, we find that Eq.~(\ref{b}) is
inconsistent on the Type I singularity solution,
so we can say that the model is free of this kind of singularity.
Nevertheless, the choice of
$g(R,G)=R$ and $f(R,G)=R$ is equivalent to
neglecting the first term of $\mathcal{F}(R,G)$, so we are considering
$\mathcal{F}(R,G)\sim R^{2}$ (strong limit when $R\rightarrow \infty$). In
this case, we find that the model could be affected by the Type I
singularity.
The physical meaning has been discussed in Sec.~V.


\end{document}